\documentclass[5p]{elsarticle}

\makeatletter
\def\ps@pprintTitle{%
	\let\@oddhead\@empty
	\let\@evenhead\@empty
	\def\@oddfoot{}%
	\let\@evenfoot\@oddfoot}
\makeatother

\usepackage{lineno,hyperref}
\usepackage{multirow} 
\usepackage{amsmath} 
\usepackage[font={scriptsize}]{subcaption} 
\usepackage{bm} 
\modulolinenumbers[1]
\graphicspath{{./figures/}} 
\journal{Nuclear Instruments and Methods in Physics Research A}

\begin{document}

\begin{frontmatter}

\title{Characterizing Subcritical Assemblies with Time of Flight Fixed by Energy Estimation Distributions }

\author[UM_address]{Mateusz Monterial\corref{mycorrespondingauthor}}
\cortext[mycorrespondingauthor]{Corresponding author}
\ead{mateuszm@umich.edu}

\author[Sandia_address]{Peter Marleau}

\author[UM_address]{Sara Pozzi}

\address[UM_address]{Department of Nuclear Engineering and Radiological Sciences, University of Michigan, Ann Arbor, MI 48109, USA}
\address[Sandia_address]{Radiation and Nuclear Detection Systems Division, Sandia National Laboratories, Livermore, CA 94551, USA}

\begin{abstract}

We present the Time of Flight Fixed by Energy Estimation (TOFFEE) as a measure of the fission chain dynamics in subcritical assemblies. TOFFEE is the time between correlated gamma rays and neutrons, subtracted by the estimated travel time of the incident neutron from its proton recoil. The measured subcritical assembly was the BeRP ball, a 4.482 kg sphere of $\alpha$-phase weapons grade plutonium metal, which came in five configurations: bare, 0.5, 1, and 1.5 in iron, and 1 in nickel closed fitting shell reflectors. We extend the measurement with MCNPX-PoliMi simulations of shells ranging up to 6 inches in thickness, and two new reflector materials: aluminum and tungsten. We also simulated the BeRP ball with different masses ranging from 1 to 8 kg. A two-region and single-region point kinetics models were used to model the behavior of the positive side of the TOFFEE distribution from 0 to 100 ns. The single region model of the bare cases gave positive linear correlations between estimated and expected neutron decay constants and leakage multiplications. The two-region model provided a way to estimate neutron multiplication for the reflected cases, which correlated positively with expected multiplication, but the nature of the correlation (sub or super linear) changed between material types. Finally, we found that the areal density of the reflector shells had a linear correlation with the integral of the two-region model fit. Therefore, we expect that with knowledge of reflector composition, one could determine the shell thickness, or vice versa. Furthermore, up to a certain amount and thickness of the reflector, the two-region model provides a way of distinguishing bare and reflected plutonium assemblies. 

\end{abstract}


\end{frontmatter}

\section{Introduction}

Multiplicity analysis has been the staple of non-destructive assay (NDA) and accountancy of fissile nuclear material for decades \cite{Ensslin1998} \cite{Hage1985} \cite{Bohnel1985} \cite{Cifarelli1986}. This NDA technique relies on the counting of single, double and higher-order multiplicities in a pre-defined time window. Fissile material has two properties that affect the multiplicity distribution: (1) multiple neutrons are emitted in coincidence from fission and (2) the subsequent fissions may produce more coincidence neutrons. Traditionally, analog coincidence circuits coupled with He-3 proportional counters are used to record this signature. 

Recent innovation in neutron coincidence counting has focused on replacing He-3 tubes altogether in favor of fast organic scintillators. Development of Fast Neutron Multiplicity Counters (FNMC) has been motivated by the promise of greater precision at lower dwell times due to intrinsically lower die-away times of these detectors leading to lower accidental coincidence rates \cite{Nakae2014}\cite{Chichester2015}. Additionally, gamma rays can be used as additional signature in pulse shape discrimination capable scintillators \cite{Enqvist2011} \cite{Miloshevsky2015}. 

Research and development is generally geared towards accounting for the differences in thermal capture He-3-based systems and FNMCs \cite{Li2016}. As a result, the development of new measurement systems is underpinned by more or less the same multiplicity analysis developed over three decades ago. Therefore, data analysis using new measurement systems continue to require both high efficiency and accurate knowledge of the efficiency of the system. This creates A design principle that drives systems toward larger sizes and geometries that limit applications and portability. However, portability is often a feature demanded by field applications such as treaty verification and nuclear emergency response. 

Instead of optimizing fast organic scintillator-based systems to the design principles of multiplicity counting, we have leveraged a few key additional signatures available to these detectors while investigating new analysis methodologies: 

\begin{enumerate}
	\item Prompt gamma rays released from fission, which are distinguished from neutrons with pulse shape discrimination (PSD) \cite{Kno2000} \cite{Adams1978}
	\item Energy imparted by incident neutrons measured through proton recoil.
	\item Timing between these detected events, measured with nanosecond and sub-nanosecond timing resolution capable of resolving the timing between individual fission events in a chain. 
\end{enumerate}

In this paper we will demonstrate how these signatures can be combined to measure Time of Flight Fixed by Energy Estimation (TOFFEE) distributions. The TOFFEE distribution is sensitive to the timing between fission events present in a measured medium. The dynamics of fission chain timing are driven by the physics of the subcritical system, mainly the multiplication of fissile material and presence of neutron moderators and reflectors. To illustrate this, we developed a two-region point kinetics model of a reflected fissile assembly. The solution to this model was then used to fit the measured TOFFEE distributions in an effort to extract physical system parameters such as multiplication and the presence of and coupling to a reflector from an inter-event timing distribution alone.

We tested our approach by fitting TOFFEE distributions from measurements of Beryllium Reflected Plutonium (BeRP) ball \cite{Mattingly2009} in a bare configuration and with iron and nickel shell reflectors. All measurements were performed with a hand-held array of eight 2''$\times$2'' Stilbene detectors. However, to draw any broader conclusions it was necessary for us to expand on the measured configurations with additional simulations. In particular, we extended the measurements by simulating thicker shell reflectors and two different materials: aluminum and tungsten. It was also necessary to test changing neutron multiplication independent of presence of reflectors by simulating bare BeRP balls with different masses. 

We found that fits to the TOFFEE distributions of the bare systems predicted the rate of change of the neutron population quite well. For each of the reflected configurations, a strong correlation between the fit and parameters and shell thickness was apparent. With some assumptions about the fissile core, we found that fits to the reflected configurations estimated the system multiplication. In addition, each reflector material had a distinct positive correlation between simulated and estimated multiplication.

\subsection{Time of Flight Fixed by Energy Estimation} \label{sec:TOFFEE}

For the past several years, Sandia National Laboratories and the University of Michigan have been collaborating on a new technique that uses both the timing between correlated gamma rays and neutrons in conjunction with the energy deposited by correlated neutrons. These signatures were first combined into Time-Correlated Pulse Height (TCPH) distributions which were shown  to be sensitive to multiplication of neutrons within fissile material and by extension the material mass and presence of intervening moderators or reflectors \cite{Miller2012, Miller2013, Monterial2013, Marleau2013, Paff2014}. The TCPH distribution is a raw representation of the measured signature in a bi-variate histogram of the neutron deposited energy and the time to correlated gamma ray. Therefore, it is difficult to interpret and challenging to model for the purpose of extracting physical parameters \cite{Monterial2017}. For this reason we re-combined the measured signatures into a single one-dimensional distribution we call the Time of Flight Fixed by Energy Estimation (TOFFEE). 

The time-of-flight (TOF) in TOFFEE is the measured time between correlated gamma rays and neutrons. In order to correct this TOF by the expected TOF of the neutron from the point of emission to the detector, we estimate its energy, $E_n$, as the energy it deposits in the detector by elastic scatter on a proton, $E_p$.  Because the neutron typically deposits only a fraction of its energy in this interaction, the estimated energy will be systematically low and thus the estimated TOF will be systematically too large. With a known source-to-detector distance $d$, it is possible to estimate the travel time difference between a neutron and gamma ray emitted simultaneously:

\begin{align}
	\label{eq:tp}
	 t_p = d \left( \sqrt{\frac{m}{2 E_p}} - \frac{1}{c} \right)
\end{align}
where $c$ is the speed of light and $m$ is neutron's mass. The calculated quantity $t_p$ is therefore the estimated difference in neutron and gamma-ray time of flight difference from the proton recoil energy. Since $E_p$ is systematically smaller than the true incident neutron energy, $t_p$ will overestimate the true time of flight difference between the neutron and gamma way. Finally, the ``fixed" in TOFFEE refers to subtracting this quantity from the measured time between gamma ray and neutron pair, $t_{n,\gamma}$. 

\begin{figure*}[h] 
 \centering
 \begin{subfigure}[h]{65mm} 
		\centering
		\includegraphics[width=\textwidth]{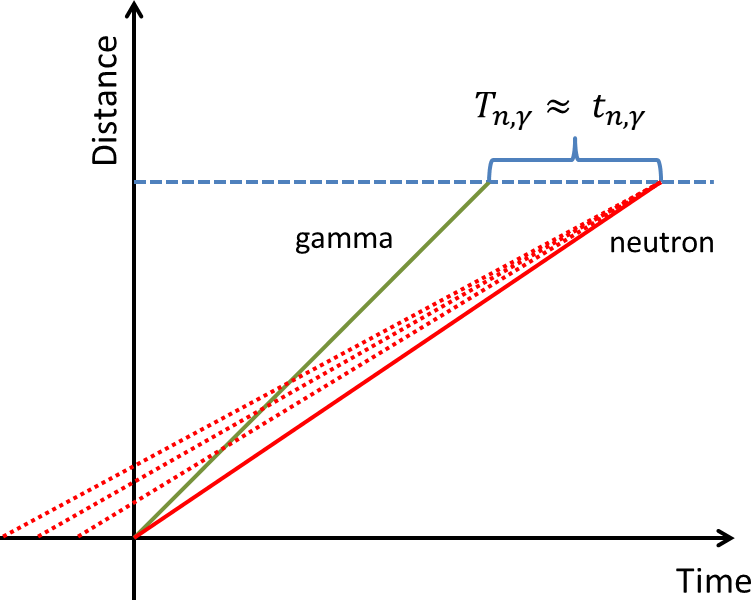}
 		\caption{Non-multiplying}
  \end{subfigure}%
     
        ~ 
  \begin{subfigure}[h]{65mm} 
		\centering
	  \includegraphics[width=\textwidth]{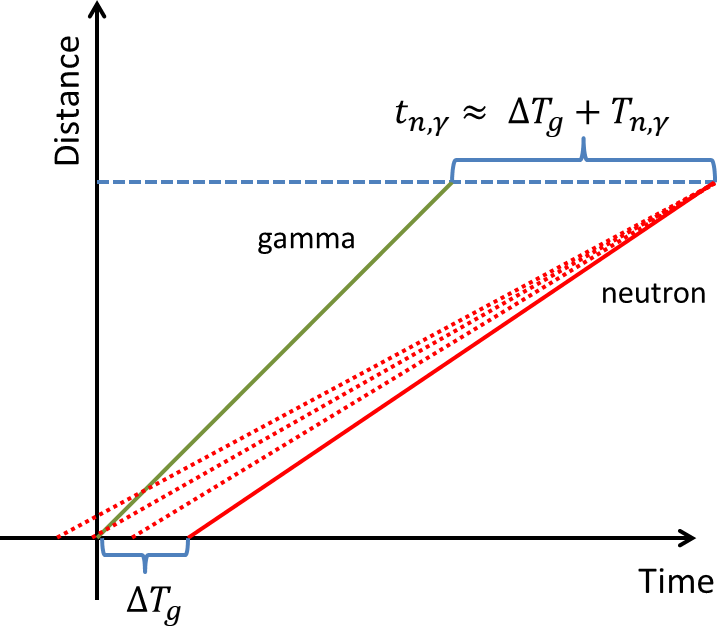}
   	\caption{Multiplying, gamma ray born first}
  \end{subfigure}%
  \begin{subfigure}[h]{65mm} 
	\centering
	\includegraphics[width=\textwidth]{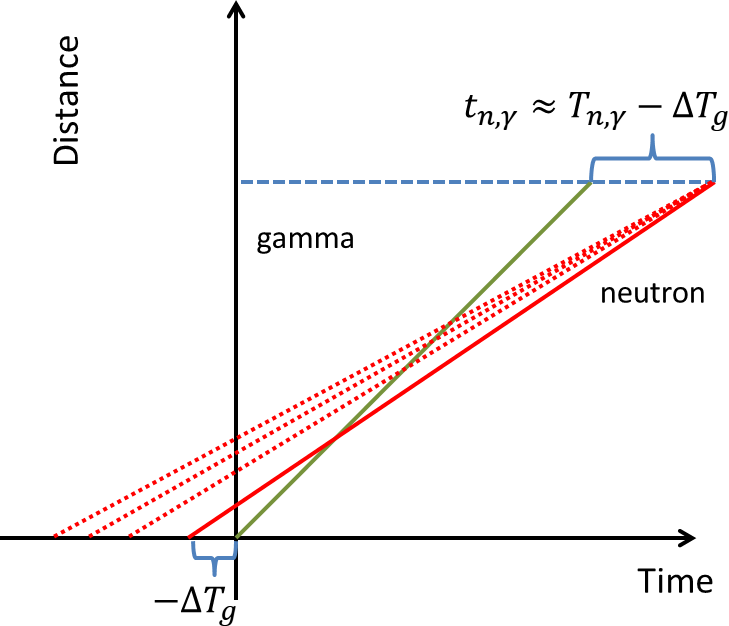}
	\caption{Multiplying, neutron born first}
   \end{subfigure}%

  \caption{Space-time diagrams of gamma ray (green) and neutron (red) particle paths from birth to detection (dashed blue line). The (a) non-multiplying  diagram depicts the simultaneous birth of particles, and the (b) and (c) multiplying diagrams depict a fission chain where each fission is separated by generation time $\Delta T_g$. The measured time-of-flight difference, $t_{n,\gamma}$, is equivalent to the true time-of-flight difference $T_{n,\gamma}$ in the non-multiplying case, but it includes the generation time in the multiplying case. The dashed red lines depict possible estimates of the neutron's velocity from proton recoil. The end-points of those dashed lines on the time-axis at the assumed source distance make up the TOFFEE distribution.}
  \label{fig:toffe_example}
\end{figure*}

For non-multiplying sources (e.g. spontaneous fission, ($\alpha$, n)), the actual travel time difference between gamma ray and neutron, $T_{n,\gamma}$, will equal the measured $t_{n,\gamma}$, as shown in Figure \ref{fig:toffe_example}(a). Therefore, TOFFEE for non-multiplying sources will be less than or equal to zero

\begin{align} \label{eq:toffee_non}
	 t_{n,\gamma} - t_p \leq 0 .
\end{align} 
In contrast, for multiplying sources the measured time difference between correlated gamma rays and neutrons will include the difference in generation time, $\Delta T_g$, as shown in Figure \ref{fig:toffe_example}(b) and (c). As a result TOFFEE for sources with present fission chains will be less than or equal to the times between fission events that gave birth to each particle

\begin{align} \label{eq:toffee_mult}
 t_{n,\gamma} - t_p \leq \Delta T_g .
\end{align}

There are three important implications from Eqs. \ref{eq:toffee_non} and \ref{eq:toffee_mult} on the relationship between TOFFEE and the type of source measured. First, there is a sharp distinction between non-multiplying and multiplying sources because the former should have a steep drop in counts on the positive side of the TOFFEE distribution. The TOFFEE distribution of a multiplying source will, in contrast, be ``smeared" in both negative and positive time directions by $\Delta T_g$.  The bi-directional smearing is exemplified in Figures \ref{fig:toffe_example} (b) and (c), and is the consequence of correlating gamma-neutron pairs where either the gamma ray or the neutron were born first. 

Second, the TOFFEE distribution is quite sensitive to the level of neutron multiplication within a source. If $k$ is defined as the ratio of the number of neutrons born in one generation to those in the previous generation, then the subcritical neutron multiplication is

\begin{align} \label{eq:mult}
 M = \frac{1}{1 - k}.
\end{align}
Neutron multiplication is equivalent to the average number of neutrons produced per starting neutron or the average length of a fission chain \cite{Serber1945}. The probability of detecting particles from the same fission grows linearly with $M$, which will be distributed according to Eq. \ref{eq:toffee_non}. Whereas, the probability of detecting particles from two different fissions in a chain increases factorially with $M$ and will be distributed according to Eq. \ref{eq:toffee_mult}. 

\begin{figure}[htbp] 
	\centering
	\includegraphics[width=85mm]{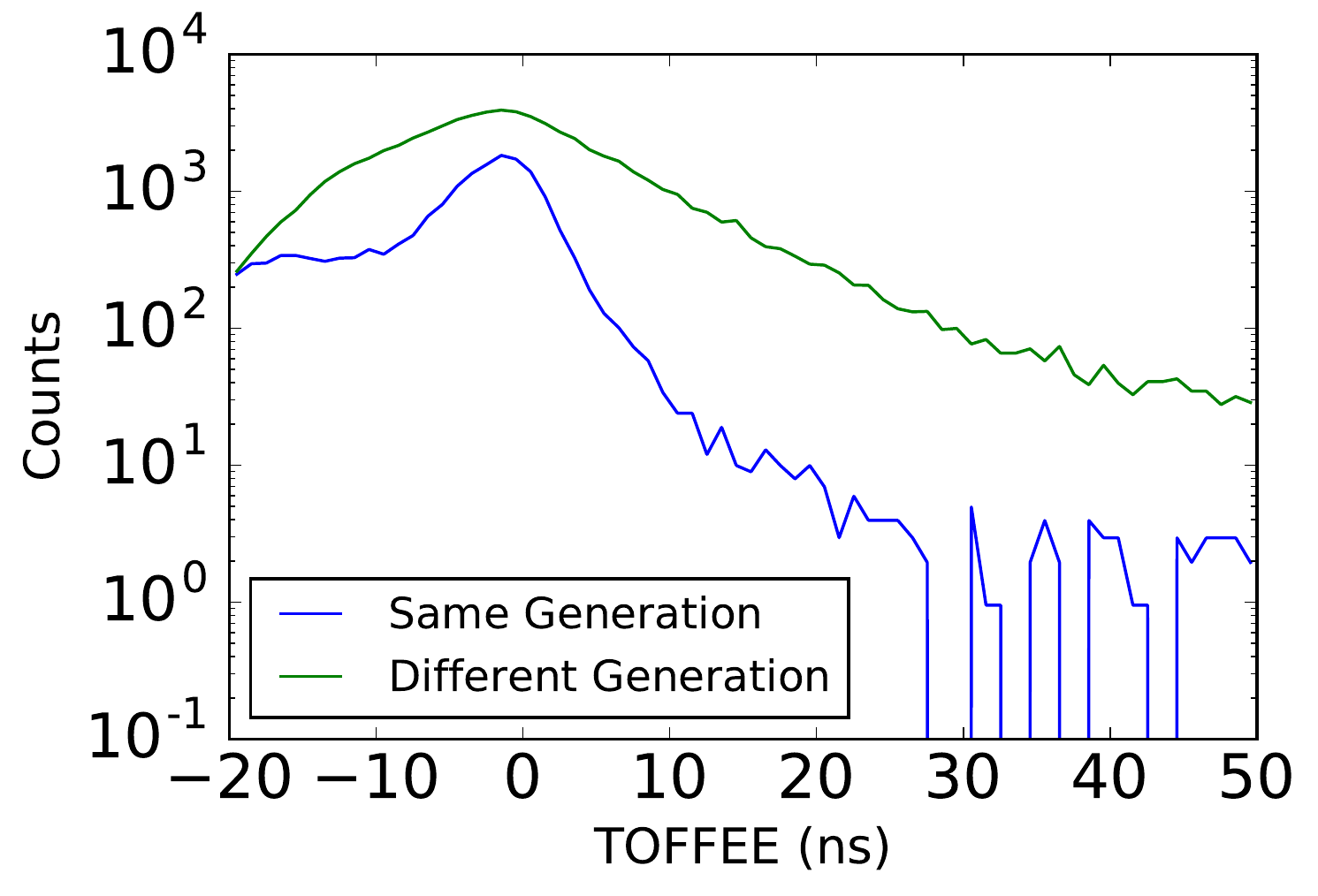}
	\caption{TOFFEE distributions of simulation of the BeRP ball constructed from gammas and neutrons originating from the same generation and different generations of fissions. There are more correlations from different generations due to neutron multiplication of the BeRP ball ($M = 4.389 \pm 0.005$).}
	\label{fig:toffee_gen_example}
\end{figure}

The contributions of the particles correlated in the same generation and different generation from a simulation of the bare BeRP ball is shown in Figure \ref{fig:toffee_gen_example}. In this example, generations are used to distinguish correlated events, because MCNPX-PoliMi output provides the generation number of a fission that originated a detected particle, but not a unique identifier of the fission event itself. Multiple fissions can belong to the same generation, because of branching in a fission chain, therefore this example is an approximation to TOFFEE distributions from same and different fissions. As a consequence, the same generation TOFFEE distribution, shown in Figure \ref{fig:toffee_gen_example}, will sometimes include the time between fissions of the same generation and will therefore also include a $\Delta T_g$ smearing term. The different generation TOFFEE distribution is not only smeared out due to the addition $\Delta T_g$, but also has noticeably more counts due to the greater probability of detecting particles that are correlated from separate fission events. 

Finally, the influence of $\Delta T_g$, as shown in Eq. \ref{eq:toffee_mult}, means that the TOFFEE distribution is simultaneously a measure of a length of a fission chain and the timing distribution of fissions within that chain. The characteristic time between fission events in a chain is indicative of the probability of fission, and the average neutron energy between fissions. In addition, for assemblies that are coupled to a reflector, the time between fission events also depends on the probability and the time for a neutron to return to the fissile material. In this paper we demonstrate that the TOFFEE distribution relates to these physical properties.

\section{Experiments}

To demonstrate the sensitivity of the TOFFEE distribution to the physical configuration of a fissile core and surrounding reflective materials, a series of measurements of the BeRP ball were made.  The BeRP ball is a 4.482 kg sphere of $\alpha$-phase weapons-grade plutonium metal (93.3 $wt\%$ Pu-239, 5.9 $wt\%$ Pu-240), originally manufactured in October 1980 by Los Alamos National Laboratory \cite{Mattingly2009}. This sphere has a mean radius of 3.7938 cm, and is encased in a 304 stainless steel shell that is 0.0305 cm thick. The measurements were conducted at the Nevada National Security Site (NNSS), with five distinct reflector configurations shown in Table \ref{tab:berp_exp}. The reflectors were made from close fitting sets hemispherical shells made of iron and nickel, with a single 4.509 cm diameter hole going through them.

\begin{table*}[htbp]
\caption{Measurement details of the various configurations of the BeRP ball with iron and nickel reflectors. The neutron multiplication was calculated from MCNP6 k-code simulation.}
\centering
\begin{tabular}{l| c c c}
\hline
\multirow{ 2}{*}{\textbf{case}} &  measurement  & rate of gamma ray      & multiplication \\ 
                                &  time (sec)   & neutron pairs (Bq) & \\
\hline
bare         & 1968  & 136 &  4.433 $\pm$ 0.001 \\ 
0.5 in Fe    & 897   & 211 &  5.584 $\pm$ 0.008 \\ 
1 in Fe      & 2095  & 280 &  6.648 $\pm$ 0.012 \\ 
1.5 in Fe    & 1497  & 239 &  7.182 $\pm$ 0.015 \\
1.0 in Ni    & 1497  & 243 &  7.472 $\pm$ 0.016 \\ \hline
\end{tabular}
\label{tab:berp_exp}
\end{table*}

In addition to a multiplying source, we measured a 21 $\mu$Ci Cf-252 source at a source to detector distance of 35 cm. This measurement was performed independently at Sandia National Laboratories.  These measurements serve to validate Monte Carlo simulations of the detection system and fissile assembly. 

\begin{figure}[h] 
	\centering
	\includegraphics[width=85mm]{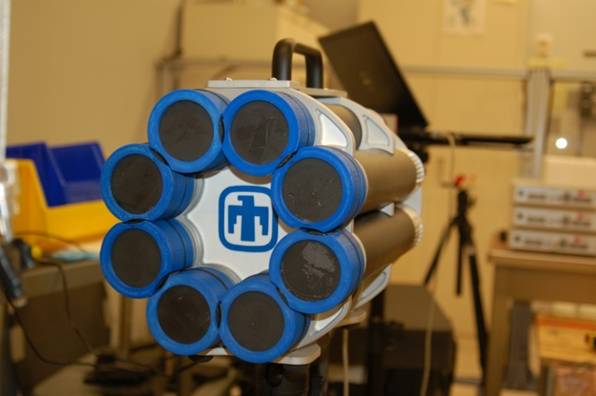}
  \caption{The front of purpose-built stilbene array used for all measurements.}
  \label{fig:stilbene_array}
\end{figure}

All measurements were performed with a purpose-built portable array of eight 2" by 2" cylindrical stilbene crystals. Each stilbene crystal was coupled to H1949-50 Hammamtsu photomultiplier tube (PMT) with a custom low voltage to high voltage bias converter. Quarter inch thick pucks of lead were attached to the front of the detectors in order to minimize count rate from uncorrelated decay gamma rays emitted by the plutonium and americium in the BeRP ball. These pucks were used for all subsequent Cf-252 and calibration experiments. A photograph of the instrument is shown in Figure \ref{fig:stilbene_array}. The anode outputs were digitized using CAEN DT5730 digitizer, capable of 14-bit vertical resolution and a 500 MHz sampling rate. The acquisition threshold was approximately 20 keVee (keV electron-equivalent), and the post-processing threshold was set to be 100 keVee. 

\subsection{Calibrations}

Energy calibration measurements were performed with a Na-22 source. Calibration constants were estimated by matching measurements to an MCNPX-PoliMi simulation \cite{Pozzi2003}. The measured pulse heights (PH) were shifted by a linear calibration formula

\begin{align}
	L = a * PH + b
\end{align}
where L is the calibrated light output and $a$ and $b$ are calibration parameters, while the simulated results were broadened by the approximate energy resolution of each detector:

\begin{align} \label{eq:resolution}
\frac{\Delta L}{L}  = \sqrt{\alpha^2 + \frac{\beta^2}{L} + \frac{\gamma^2}{L^2}}
\end{align}
where $\alpha$, $\beta$ and $\gamma$ parameters include contributions from light transmission within detector cell, statistical fluctuations of light production and electronic noise, respectively \cite{Scholermann1980}. 

\begin{figure}[h] 
  \centering
	\includegraphics[width=85mm]{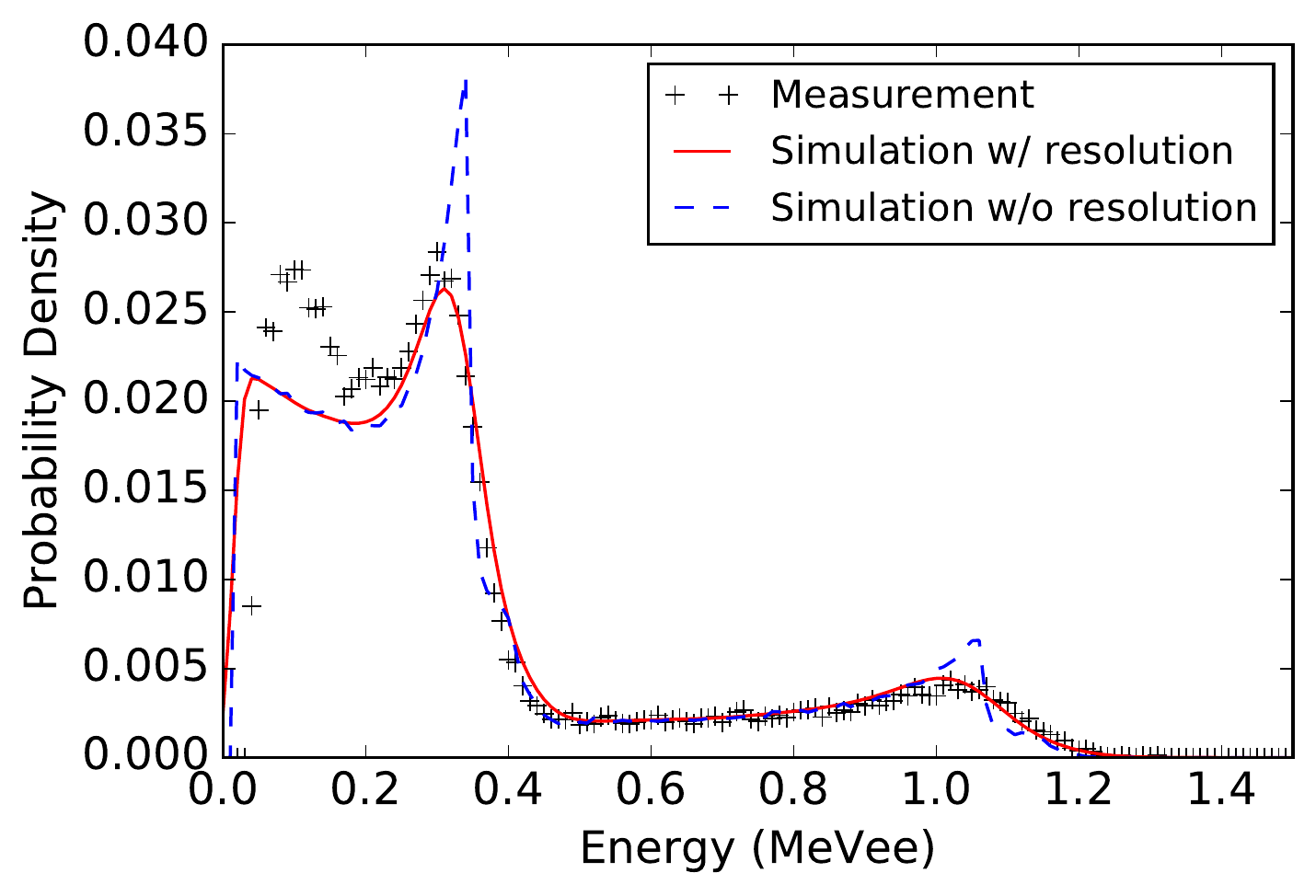}
 	\caption{Measured and simulated spectra of Na-22 source matched with optimum resolution and calibration parameters.}
  \label{fig:phd_sim_msr}
\end{figure}

Levenberg-Marquardt algorithm was employed to find the optimum calibration and resolution parameters and an example of the results are shown in Figure \ref{fig:phd_sim_msr}. Extra weight was given to the regions of the spectra around the Compton edges, and the back-scatter peak was ignored. 

The neutron light output yield for 2" stilbene crystals was measured by Bourne et al. in a separate set of experiments \cite{Bourne2017}. The set of proton recoil energies and corresponding calibrated light outputs were fitted to Birks' formula 

\begin{align} \label{eq:birks}
L(E_p) = \int \frac{a}{1 + b (dE/dx)} dE
\end{align}
where $dE/dx$ is the proton stopping power in stilbene \cite{Norsworthy2017}. The fitted parameters of $a$ and $b$ were found to be 1.63 (MeVee/MeV) and 27.83 (mg/(cm$^2$ MeV)), respectively. The integrand in Eq. \ref{eq:birks} was evaluated for deposited energies ranging from 1 keV to 250 MeV, and the results were saved in a lookup table. Linear interpolation of the values in the table was used to calculate light output from simulations and approximate proton recoil energy from light output in measurement. 

We employed a Bayesian approach to pulse shape discrimination (PSD) \cite{Monterial2015}, in order to separate neutrons from gamma rays and give each prospective particle detection an appropriate weight. This approach allowed us to study the effect PSD cuts have on the outcome of the final TOFFEE distribution. As expected the ``harder" PSD cuts are akin to raising the energy threshold, due to the overlap between neutrons and gamma rays PSD at lower energies. We used the time differences between 90\% and 10\% of the pulse cumulative trapezoidal integral as the PSD parameter. 

\section{Simulation Validations}

The measured configurations of the BeRP ball, shown in Table \ref{tab:berp_exp}, include three sets of shell thickness and two types of shielding material. We used simulations to expand the range of shell thicknesses, explore other reflector materials, and vary the mass of the BeRP ball by changing the diameter of the sphere. The simulations of the TOFFEE distributions were performed with MCNPX-PoliMi, and details simulated materials are provided in Table \ref{tab:sim}. The k-code calulcations performed with MCNP6 used the same materials, and $k_{eff} $ was estimated over 50 cycles with 1 million neutrons per cycle. In order to have confidence in these results it was necessary to validate the simulations by comparing them with measurements.

\begin{table*}[htbp]
	\caption{Specifications for the materials used in simulation of the BeRP ball with various reflectors.}
	\centering
	\begin{tabular}{l| c c c}
		\hline
		\multirow{2}{*}{material}   &  isotopic concentration                                         	& cross-section 	& density \\ 
		                   &  mass fraction (wt\%)  											& library name			& (g/cc) \\
		\hline
		Bare (Pu)     &   93.27 $^{239}$Pu, 5.91 $^{240}$Pu                          & endf70j    & 19.6 \\
	                       &   0.45 $^{238}$U, 0.25 $^{241}$Pu                       		&       &   \\ \hline
		Nickel          &   67.20 $^{58}$Ni, 26.78 $^{60}$Ni, 3.84 $^{62}$Ni   & rmccs    &   8.909 \\ 
		                   &   1.18 $^{61}$Ni, 1.01 $^{64}$Ni                                   &              &    \\ \hline
		Iron             &   91.90 $^{56}$Fe, 5.65 $^{54}$Fe 							 & rmccs     &    7.874 \\ 
		   		           &   2.16 $^{57}$Fe, 0.29 $^{58}$Fe                                 &    &      \\  \hline
		Aluminum    &   100 $^{27}$Al			    									                                                         & endf71x    &  2.7 \\ \hline
		Tungsten    &   30.69 $^{184}$W, 28.79 $^{186}$W, 26.26 $^{182}$W  & rmccs	 &  19.3 \\ 
						    &   14.26 $^{183}$W, 0.12 $^{180}$W                                &  	 &    \\ \hline
	\end{tabular}
	\label{tab:sim}
\end{table*}

\subsection{Cf-252}

\begin{figure}[htbp] 
	\centering
	\begin{subfigure}[h]{85mm} 
		\centering
		\includegraphics[width=\textwidth]{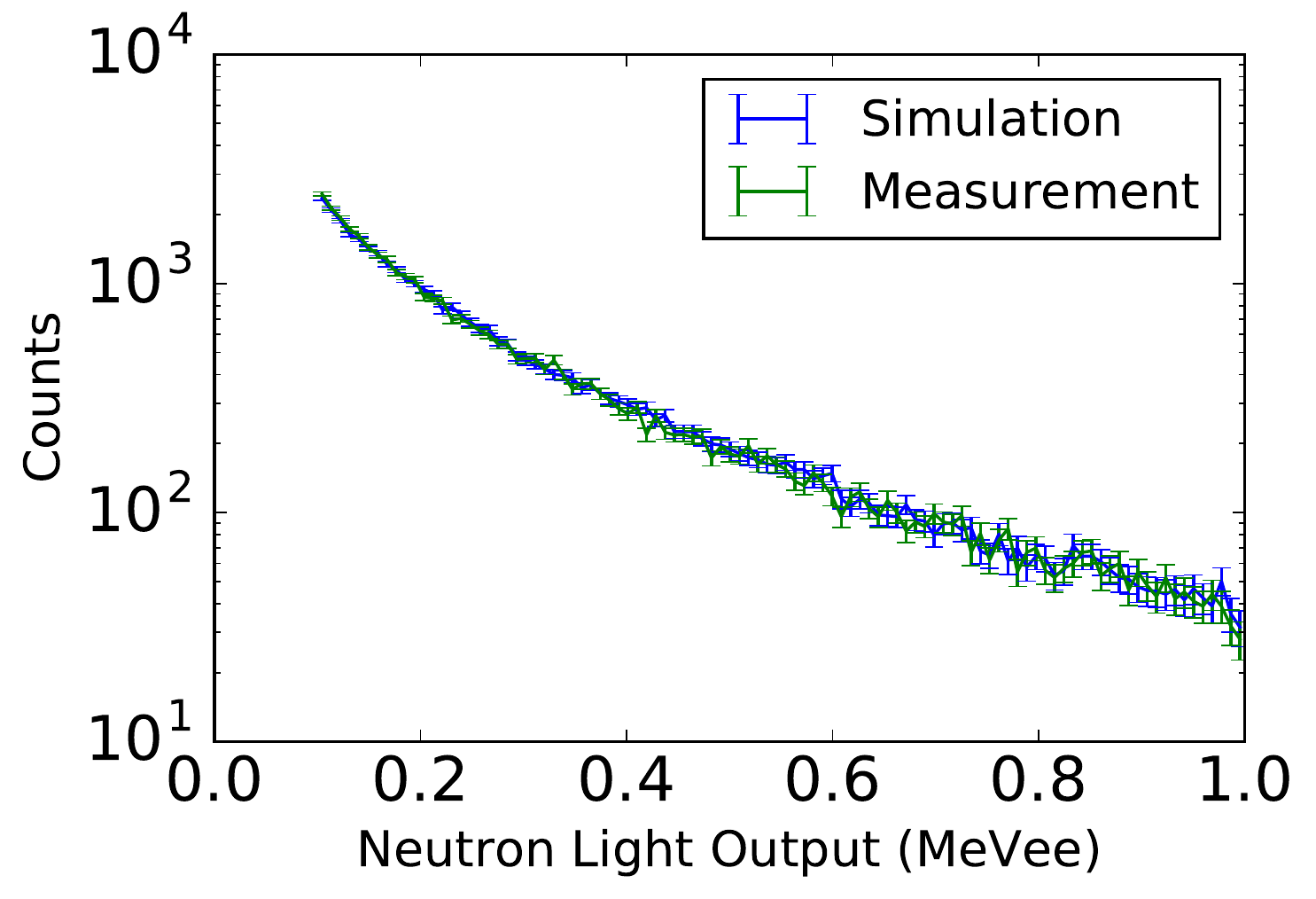}
		\caption{}
	\end{subfigure}%
	
	~ 
	\begin{subfigure}[h]{85mm} 
		\centering
		\includegraphics[width=\textwidth]{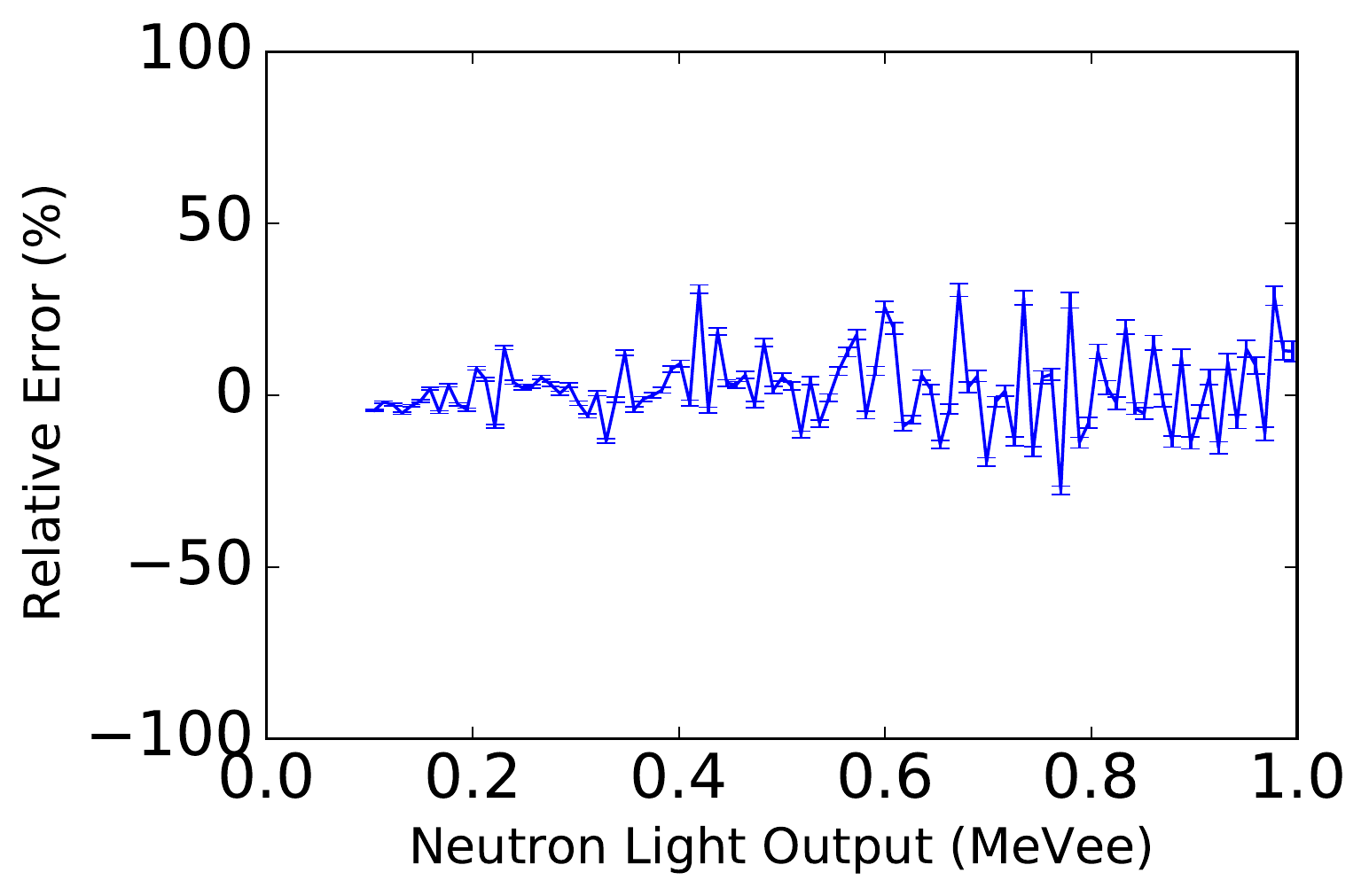}
		\caption{}
	\end{subfigure}%
	
	\centering
	\caption{Measurement and simulation comparison of the Cf-252 source (a) pulse height distribution of gamma ray correlated neutrons and (b) corresponding relative error of the simulation.}
	\label{fig:cf252_phd}
\end{figure}

First, we compared the neutron pulse height distributions (PHDs), which should test the energy calibration and neutron light output function. We limited the neutrons to those that were correlated with gamma rays inside a 2 $\mu$s window. The measured and simulated PHDs shown in Figure \ref{fig:cf252_phd} overlap with the entire range of measured energies, without any noticeable systematic bias and within the statistical error. The statistical fluctuations are reflected in the relative error, which oscillates around zero. 

\begin{figure}[htbp] 
	\centering
	\begin{subfigure}[h]{85mm} 
		\centering
		\includegraphics[width=\textwidth]{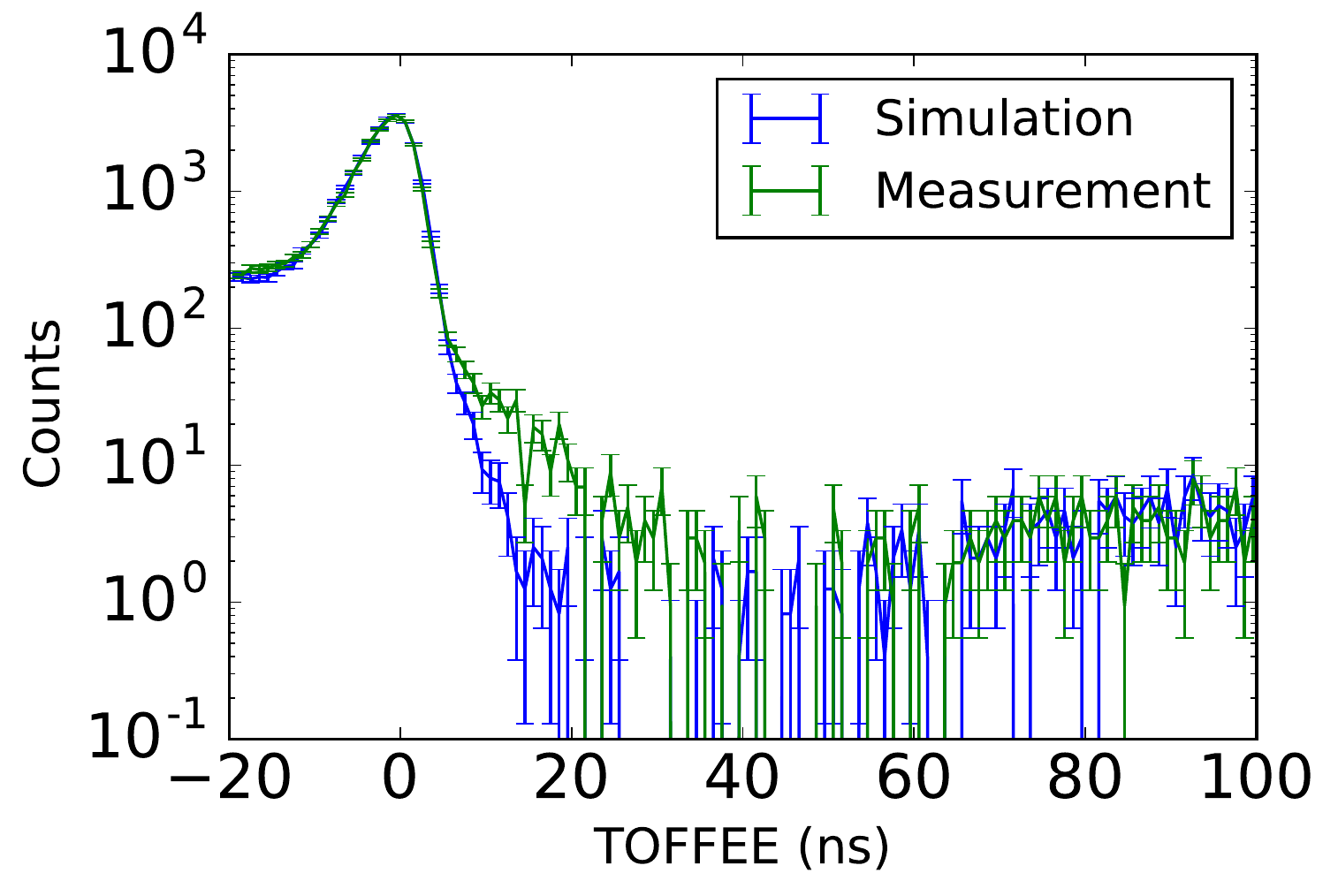}
		\caption{}
	\end{subfigure}%
	
	~ 
	\begin{subfigure}[h]{85mm} 
		\centering
		\includegraphics[width=\textwidth]{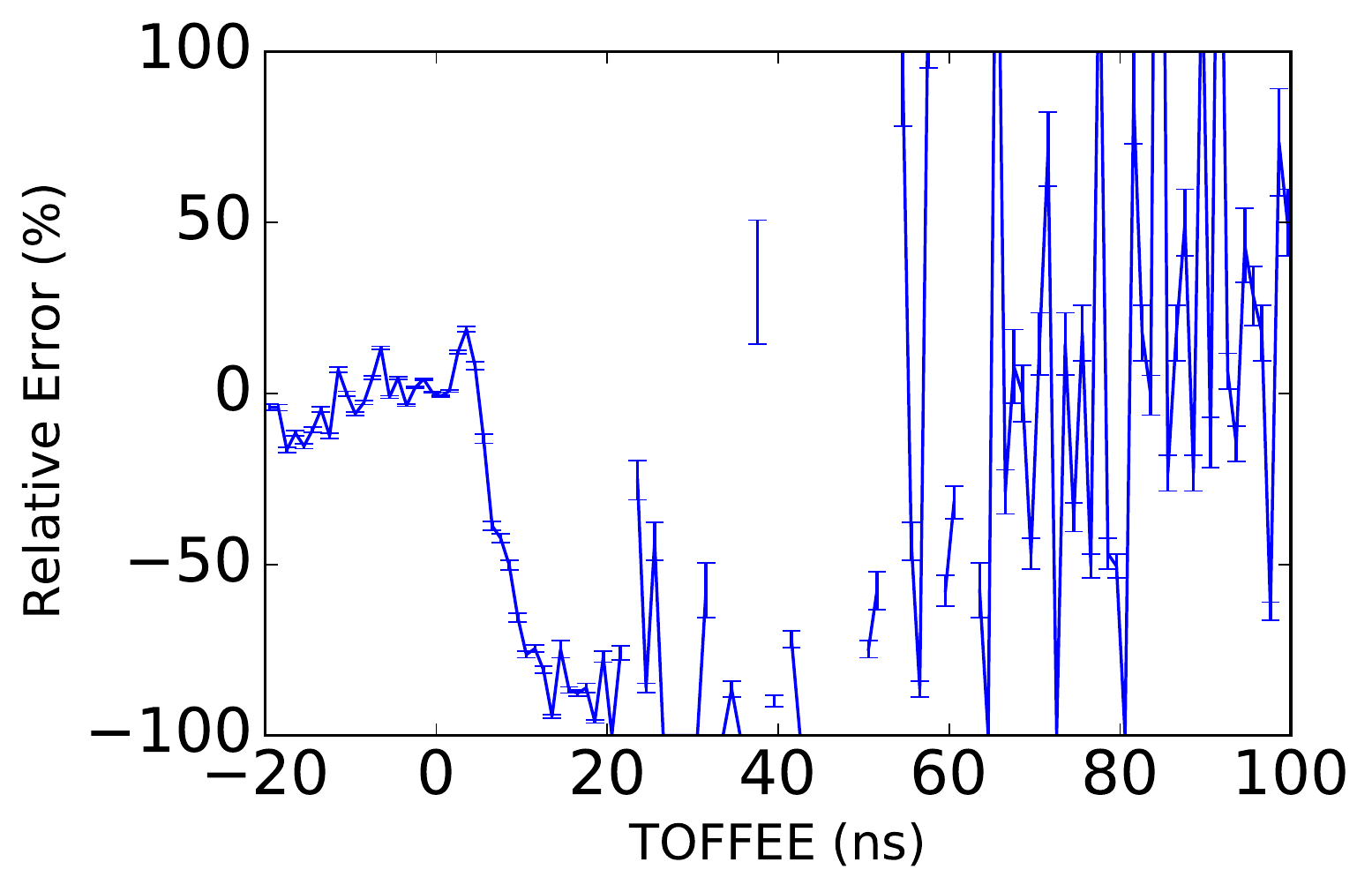}
		\caption{}
	\end{subfigure}%
	
	\centering
	\caption{Measurement and simulation comparison of the Cf-252 source (a) TOFFEE distribution and (b) corresponding relative error of the simulation.}
	\label{fig:cf252_toffee}
\end{figure}

In contrast to the PHD, which is relatively featureless, the TOFFEE distributions shown in Figure \ref{fig:cf252_toffee} have several features whose shape depend on the detector response. The most significant feature is the bell-like curve between -10 and 5 ns which includes the vast majority of correlated counts. The width of these curves line up with each other, indicating that the energy calibration and corresponding thresholds are matched, and that time resolution is properly applied. In addition, the width is affected by the source-to-detector distance, which in both the simulation and measurement was 35 cm. 

The higher counts in the measurement in the region between -20 and -10 ns is partly due to PSD misclassification, where gamma-gamma correlations are mistakenly classified as gamma-neutron. There is also good agreement in the region beyond 60 ns, where the effects of scattering from the floor is evident. There are not many counts in that region, which contributes to the erratic relative error, but the simulation and measurement match within the statistical error. The region of the largest notable error lies roughly between 5 and 25 ns, right around the steep drop in counts expected from a non-multiplying source. 

Finally, there is the rate of ``accidental" correlations that depend on the source strength and appear as a flat background in the TOFFEE distribution. The contribution from accidentals is estimated by averaging counts in each bin of a region offset by 1000 to 1500 ns from each coincidence trigger. This is then subtracted from the TOFFEE distribution. In simulation this can be varied by manipulating the effective equivalent ``measurement" time. 

\subsection{BeRP Ball} \label{sec:berp_ball}

The BeRP ball is a much more complicated source compared to Cf-252. It's a distributed spherical source having a diameter of 7.59 cm and a multiplication of 4.4, and therefore cannot be treated as a non-multiplying source. For our simulation we evenly distributed spontaneous fissions of Pu-240 and ignored the more complicated mix of isotopes that become ingrown over time. 

\begin{figure}[h] 
	\centering
	\begin{subfigure}[h]{85mm} 
		\centering
		\includegraphics[width=\textwidth]{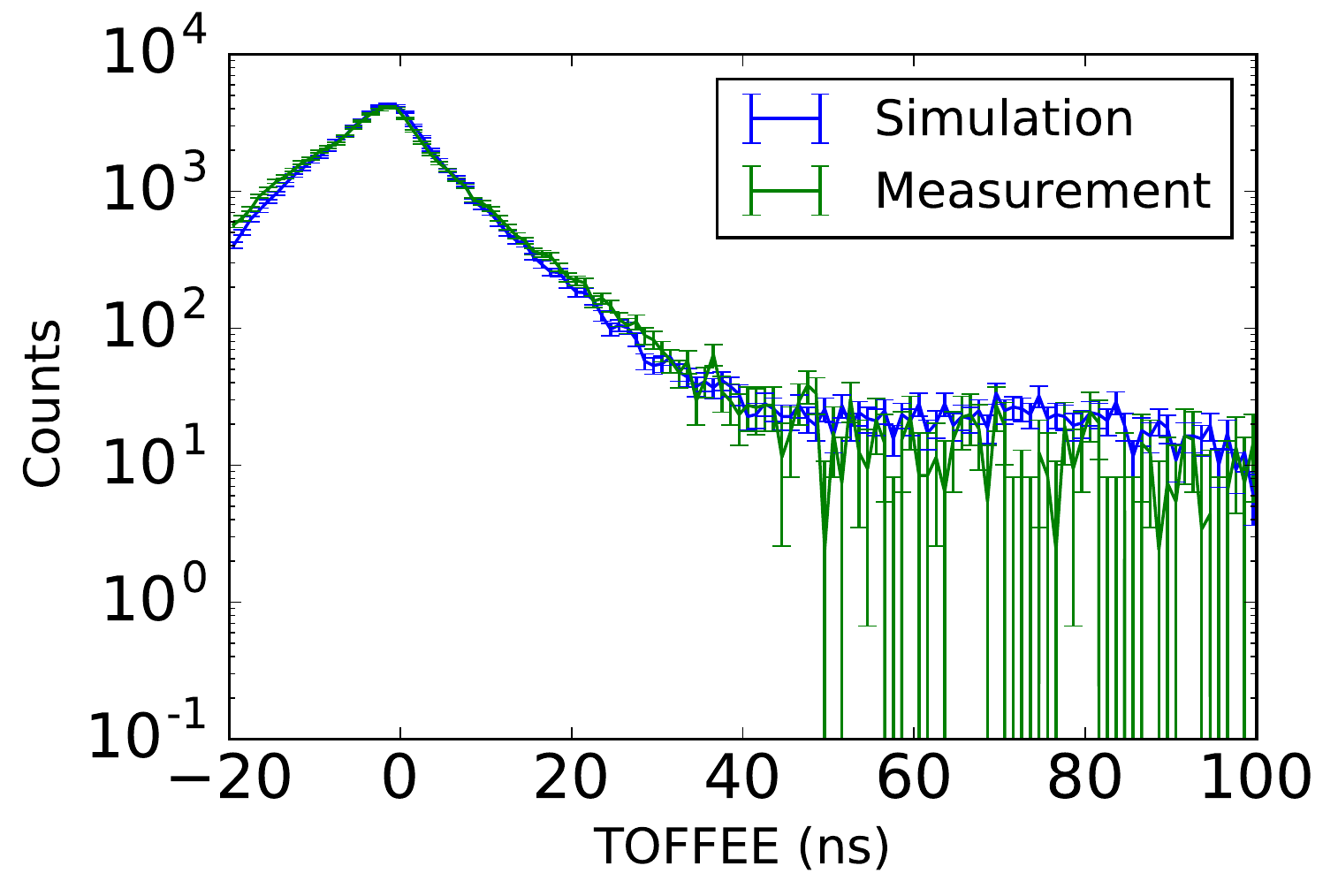}
		\caption{}
	\end{subfigure}%
	
	~ 
	\begin{subfigure}[h]{85mm} 
		\centering
		\includegraphics[width=\textwidth]{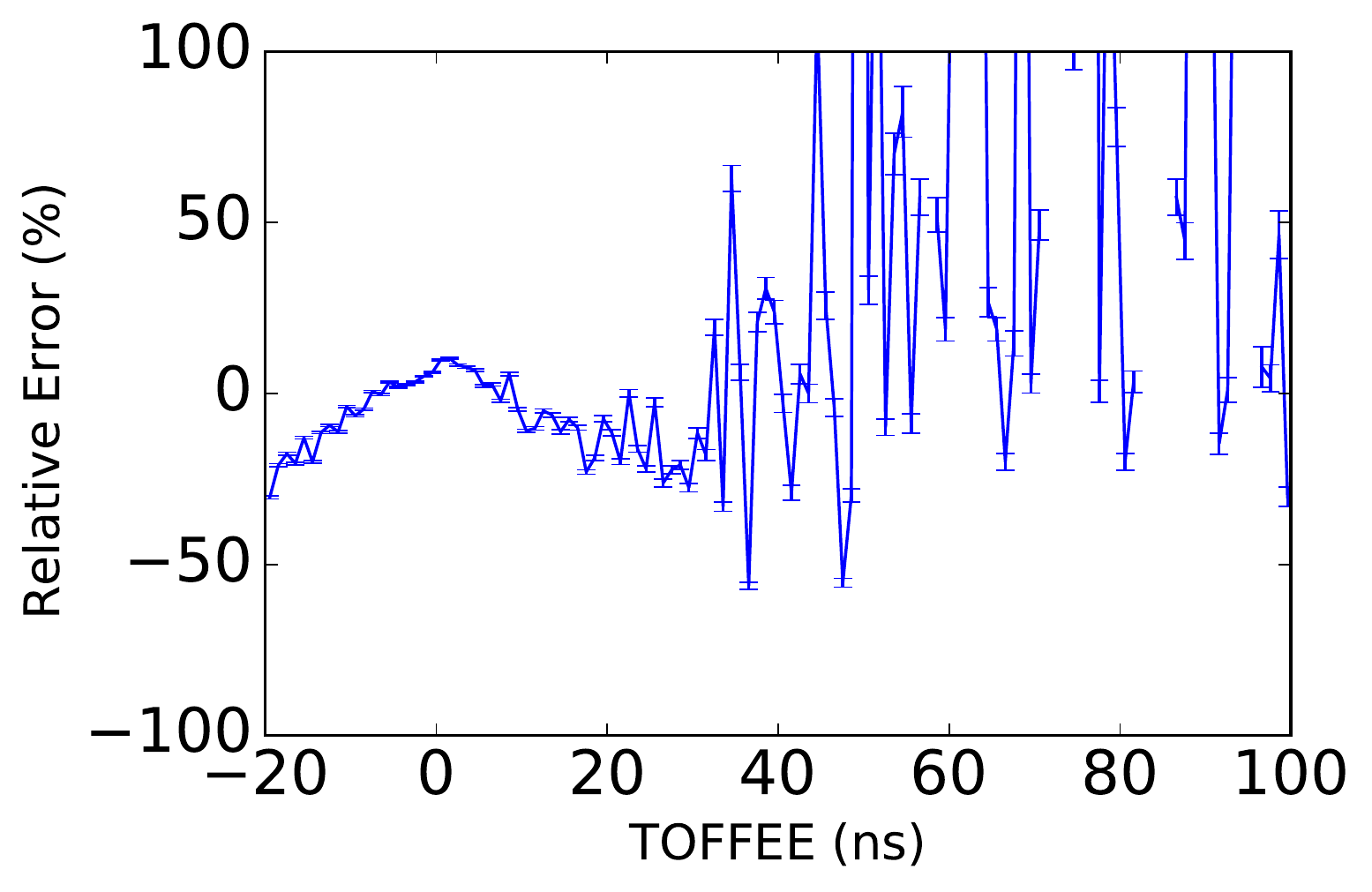}
		\caption{}
	\end{subfigure}%
	
	\centering
	\caption{Measurement and simulation comparison of the bare BeRP ball (a) TOFFEE distribution and (b) corresponding relative error of the simulation.}
	\label{fig:bare_toffee}
\end{figure}

The bare configuration comparison, shown in Figure \ref{fig:bare_toffee}, shows that the measured TOFFEE distribution is just slightly wider. The larger source of discrepancy is in the region between 40 and 100 ns, which is dominated by reflection from the floor. There are many time bins that are within statistical agreement in that region, but also a handful that have no counts at all. The problem is that the accidental background rate is much lower in the simulation (1 per ns) compared to the measurement (62 per ns) because of the lack of ingrown isotope sources in the former. The higher accidental background competes with the effect of room return and is statistically significant when the two are subtracted, which is apparent from the large uncertainties. 

\begin{figure}[h] 
	\centering
	\begin{subfigure}[h]{85mm} 
		\centering
		\includegraphics[width=\textwidth]{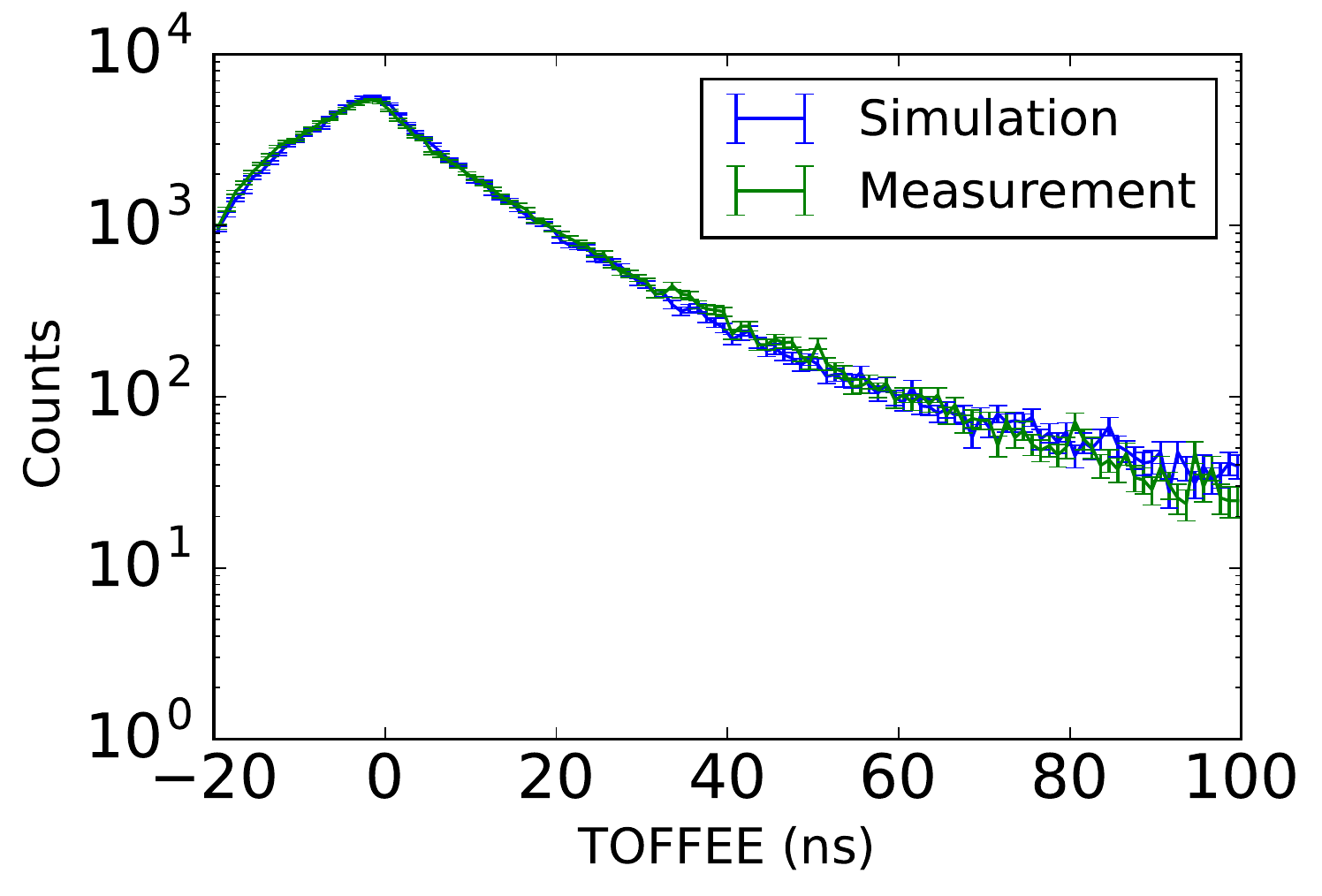}
		\caption{}
	\end{subfigure}%
	
	~ 
	\begin{subfigure}[h]{85mm} 
		\centering
		\includegraphics[width=\textwidth]{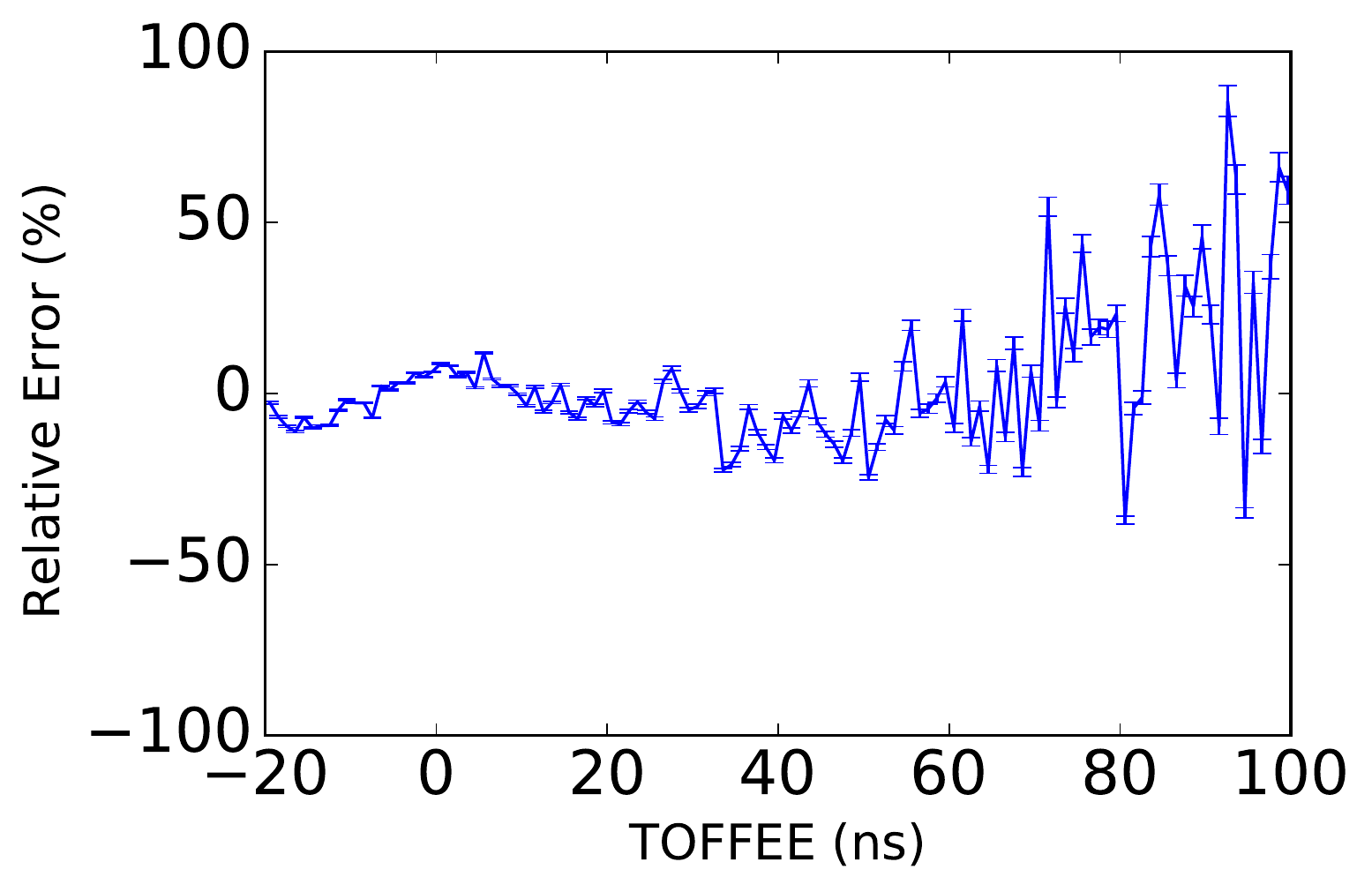}
		\caption{}
	\end{subfigure}%
	
	\centering
	\caption{Measurement and simulation comparison of the BeRP ball with 1 inch iron shielding (a) TOFFEE distribution and (b) corresponding relative error of the simulation.}
	\label{fig:fe_toffee}
\end{figure}

We found that the overall agreement between simulated and measured TOFFEE distributions improves as shielding material is added on. The BeRP ball with 1 inch iron is shown in Figure \ref{fig:fe_toffee} as a representative example of the improvement. It appears that the time smearing associated with longer fission chains sweeps up some discrepancies caused by room return, and accidental background. 

\section{Exponential Fitting} 

\subsection{Two-region Point Kinetics} \label{sec:two_region}

In this work, TOFFEE distributions were characterized by fitting the time range between zero and 100 ns with a double exponential function. We first motivate the double exponential distribution as a plausible physical response of reflected fissile material. As described in Section \ref{sec:TOFFEE}, the spread in the TOFFEE distribution of a multiplying source is driven by the generation time, $\Delta T_g$, between the detected gamma rays and neutrons. The probability of detecting these particles is governed by the time dependent population of fissions, or the corresponding neutrons that propagate fission chains. Point kinetics equations are a well established method for studying the time-dependent neutron populations in a nuclear reactor. However, modeling neutron behavior inside reflected assemblies required a two-region kinetic model \cite{Cohn1962, Spriggs1997}.

In this work we deviate from reactor point kinetics by ignoring delayed neutron precursors that result from the cascade of decays of fission product isotopes since the correlation times under consideration are on the order of only a hundred nanoseconds \cite{Dude62}. These isotopes are typically organized into six groups with half-lives ranging from hundreds of milliseconds to tens of seconds. In this work we ignore the source of these delayed neutrons, since the TOFFEE correlation window of interest is only on the order of a hundred nanoseconds. Ignoring delayed precursors, the time-dependent neutron population of prompt neutrons in a reflected assembly can be approximated by:

\begin{align} \label{eq:dNc}
\frac{dN_c}{dt} &= \frac{k_c-1}{l_c} N_c + f_{rc} \frac{N_r}{l_r} \\
\frac{dN_r}{dt} &= f_{cr} \frac{N_c}{l_c} - \frac{N_r}{l_r} \label{eq:dNr}
\end{align} 
where: 

\begin{itemize}
	\setlength\itemsep{-0.5em}
	\item[] $N_c$ is the number of neutrons in the fissile core region
	\item[] $N_r$ is the number of neutrons in the reflector 
	\item[] $k_c$ is the multiplication factor in the fissile core region
	\item[] $l_c$ is the neutron lifetime in the fissile core region
	\item[] $l_r$ is the neutron lifetime in the reflector region
	\item[] $f_{cr}$ is the fraction of neutrons that leak from the fissile core region into the reflector
	\item[] $f_{rc}$ is the fraction of neutrons that leak from the reflector back into the core
\end{itemize}
Note that the $k_c$ is different from the multiplication factor $k$ in Eq. \ref{eq:mult}. The former is the property of only the core, while the latter is the property of the whole \textit{system} (i.e. the core and reflector assembly).

The system of equations in Eqs. \ref{eq:dNc} and \ref{eq:dNr} can be solved by converting them to a second order differential equation:

\begin{align} \label{eq:second_order}
	l_r l_c \frac{d^2N_c}{dt^2} + (l_c - l_r(k_c - 1)) \frac{dN_c}{dt} - (f + k_c - 1) N_c = 0 
\end{align}
The new variable $f$ is the fraction of neutrons that leak out of the core and are reflected back, which is just the product of two previously defined terms
\begin{align}
f &= f_{rc}f_{cr}
\end{align} \label{eq:f}

In order to fully solve this problem, we enforce two initial conditions:

\begin{align}
N_c(0) &= N_o \\
N_r(0) &= 0
\end{align} 
at $t = 0$ the neutron population in the core is $N_o$ and no neutrons are present in the reflector.  The solution to Eq. \ref{eq:second_order}, given these initial conditions is a familiar double exponential:

\begin{align} \label{eq:Nc}
N_c(t) = N_o \left[ (1 -R) e^{t r_1} + R e^{t r_2} \right]
\end{align}
where the roots to the characteristic polynomial are

\begin{align}
r_1 =  \frac{-\sqrt{4 l_c l_r (f + k_c-1) + (l_c - l_r (k_c - 1))^2} - l_c + l_r (k_c - 1)}{2 l_c l_r} \\
r_2 =   \frac{\sqrt{4 l_c l_r (f + k_c-1) + (l_c - l_r (k_c - 1))^2} - l_c + l_r (k_c - 1)}{2 l_c l_r}
\end{align}
and scaling ratio $R$ is 
\begin{align} \label{eq:R}
R = \frac{r_1 - \alpha}{r_1 - r_2}
\end{align} 
where
\begin{align} \label{eq:alpha}
\alpha &= \frac{k_c - 1}{l_c} \\
\end{align} 
$f$ and $k_c$ are constrained to be less than 1. We have found that R falls within the range 0 to 1 for all plausible combinations of these variables. 

\subsection{Single-region Point Kinetics}

The parameter $\alpha$ from Eq. \ref{eq:alpha} represents the rate of loss of neutrons in a bare system (i.e. $f=0$) in which case the time dependent neutron population is simply

\begin{align} \label{eq:Nc_single}
N_c(t) = N_o e^{\alpha t}.
\end{align}
This solution to the neutron population behaviour in an unreflected assembly is a starting point to Rossi-alpha analysis \cite{Orndoff1957}. We will use it to fit the TOFFEE distributions of BeRP balls with various fissile material masses.

\section{Results and Discussion}

\subsection{Bare Configurations}

Neutron multiplication and reflector thickness are correlated, since neutron reflection increases the neutron population and average length of fission chains. In order to study the effect of multiplication independently of the effects of reflector,we simulated bare BeRP balls with various masses, ranging from 1 to 8 kg. Because of the lack of reflection, we fit the resultant TOFFEE distributions to Eq. \ref{eq:Nc_single}. The fits are made to the positive side of the TOFFEE distribution ranging from 0 to 100 ns. 

\begin{figure}[htbp] 
	\centering
	\includegraphics[width=85mm]{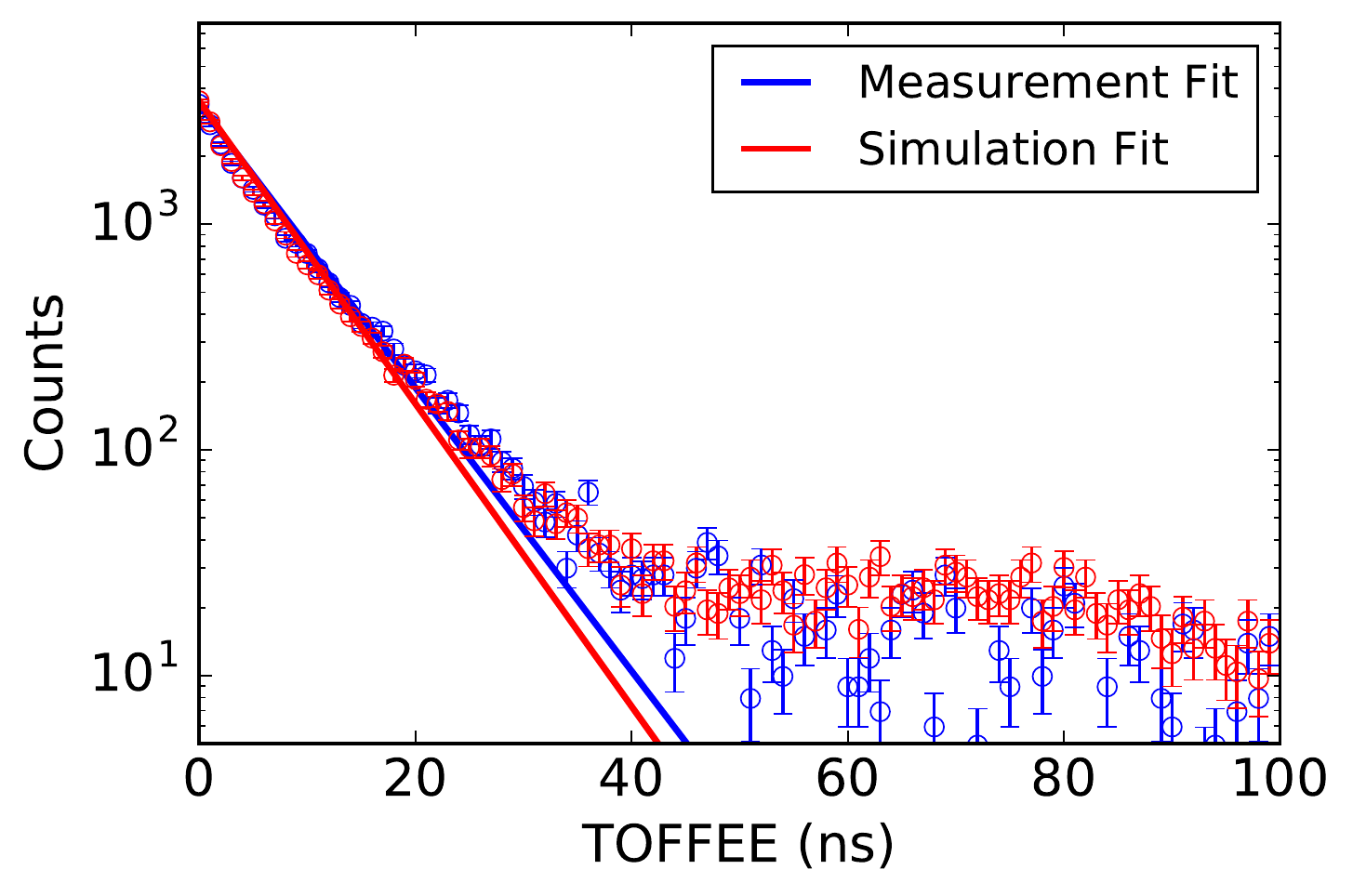}
	\caption{Comparison of the measured and simulated bare BeRP ball TOFFEE distributions and exponential fits from Eq. \ref{eq:Nc_single}.}
	\label{fig:bare_exp}
\end{figure}

A comparison of the measured and simulated BeRP ball is shown in Figure \ref{fig:bare_exp}. As explained in Section \ref{sec:berp_ball}, there is some disagreement at later times due to competing effects of floor reflection and accidental correlations. However, the fits and resulting $\alpha$ parameters for measurement ($0.144 \pm 0.003$) and simulation ($0.153 \pm 0.004$) are within two standard deviations of each other.  

\begin{figure}[htbp] 
	\centering
	\includegraphics[width=85mm]{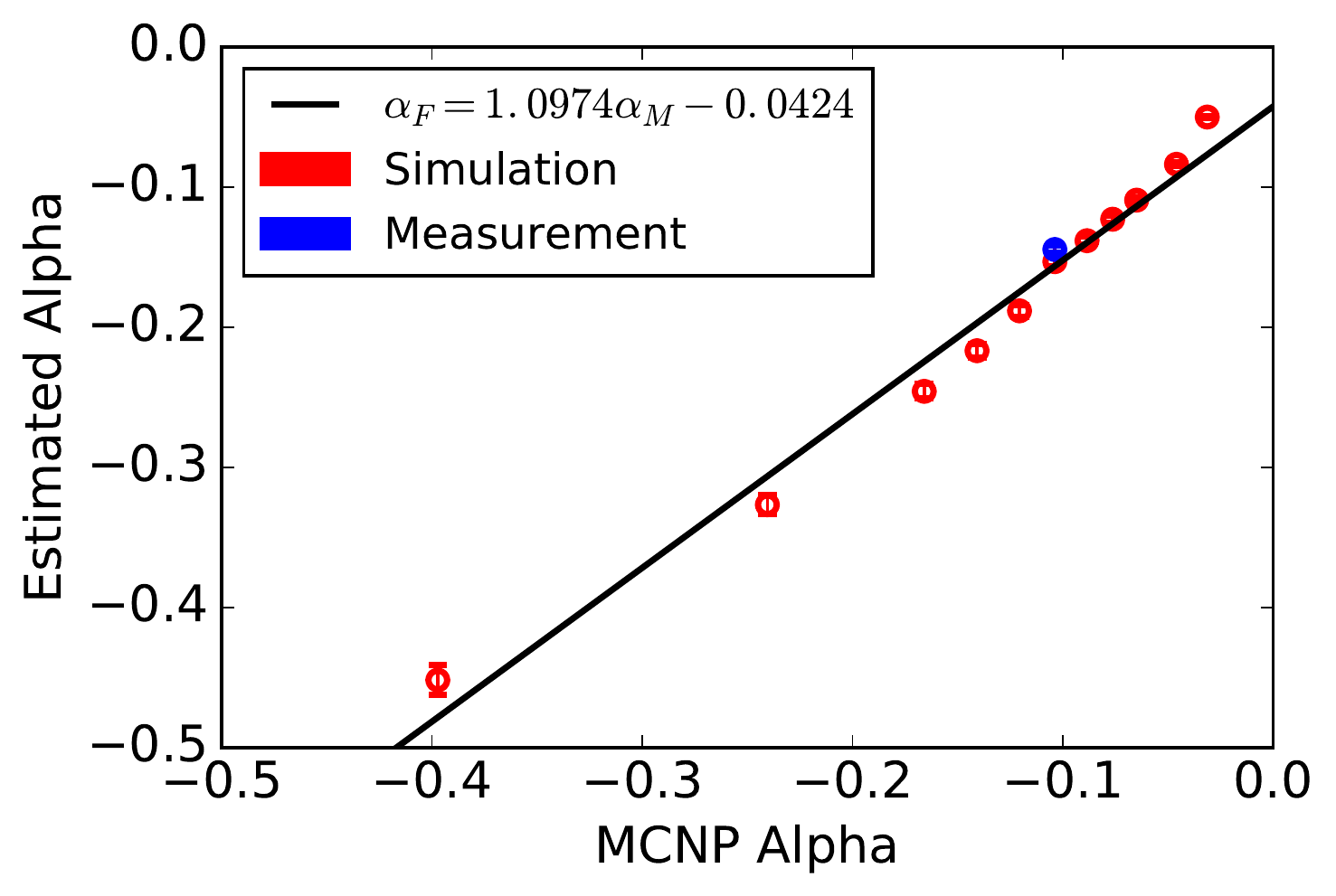}
	\caption{The fitted ($\alpha_F$) and calculated, from MCNP6, ($\alpha_M$) alpha parameters for BeRP balls with mass ranging from 1 to 8 kg. A linear regression was performed with the resulting relationship shown in the legend and a correlation coefficient of 0.9890.}
	\label{fig:bare_alpha}
\end{figure}

The multiplication factors and neutron lifetimes for the bare cases were tallied in MCNP6 simulations, and Eq. \ref{eq:alpha} was used to calculate corresponding $\alpha$ parameters. Figure \ref{fig:bare_alpha} shows the comparison of these MCNP derived alpha values with the alpha values derived from the exponential fits. The relationship between estimated and MCNP alphas is linear with a correlation coefficient greater than 0.98, and the slope of 1.0974. 

As shown in Figure \ref{fig:bare_alpha}, there is a slight deviation from the linear trend for the actual BeRP ball simulation and measurement. For all other masses we removed the thin stainless steal shell and simulated a truly bare Pu sphere. 

\begin{figure}[htbp] 
	\centering
	\includegraphics[width=85mm]{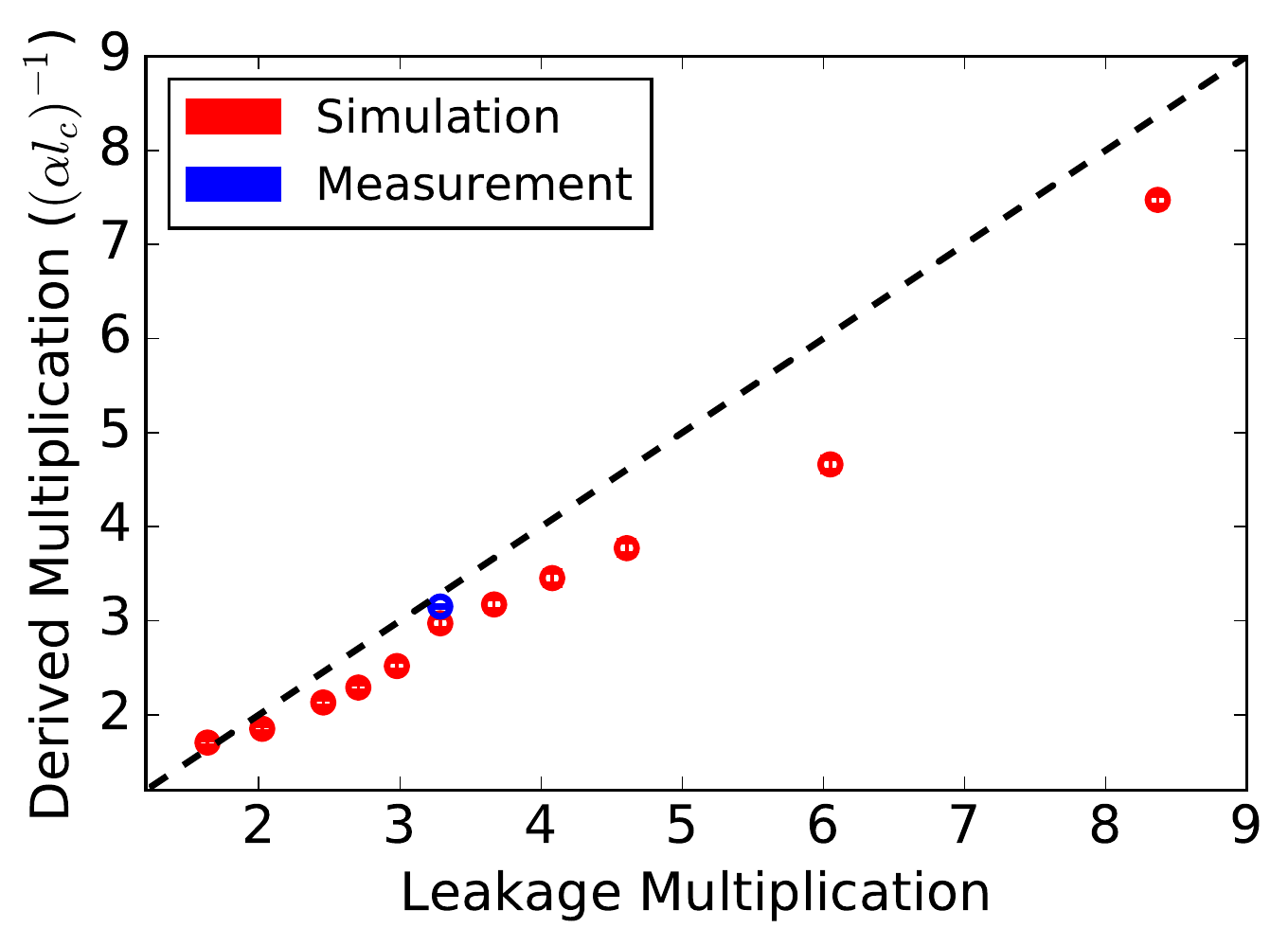}
	\caption{Derived neutron multiplications from TOFFEE fits of the bare BeRP balls with different masses with the corresponding leakage multiplications obtained through MCNP6 simulations. The dashed line corresponds to perfect agreement between derived and actual leakage multiplication, with the points above and below corresponding to overestimation and underestimation, respectively}
	\label{fig:bare_leak}
\end{figure}

Neutron multiplication can be derived from neutron decay constant and core lifetime by rearranging Eq \ref{eq:alpha}:

\begin{align}
M = -\frac{1}{\alpha l_c}.
\end{align}
We derived neutron multiplications from fitted alpha parameters by using previously tallied core neutron lifetimes from MCNP6. As expected there was a positive linear correlation between the derived and actual neutron multiplications, with the derived values that underestimate the neutron multiplication obtained from MCNP6 k-code calculations. We found that our derived multiplications better aligned with leakage multiplication, as shown in Figure \ref{fig:bare_leak}, with an average relative error deviation of 10.6\%. 

Neutron leakage is a product of the neutron multiplication and probability of neutron escape from the core. The better agreement makes some qualitative sense because neutrons that leak are the only ones that are available for detection. Furthermore, due to self-shielding the gamma rays available for detection are predominately drawn from the surface of the BeRP ball. As a result we disproportionately detect from correlations of particles originating from fission chains near the surface, which due to neutron leakage will have a shorter length than the average.  

\subsection{Multiplication} \label{sec:multiplication}

Next we moved to fitting Eq. \ref{eq:Nc}, derived from two-region point kinetics model in Section \ref{sec:two_region}, to the reflected BeRP cases. We first used the alpha parameter determined from the fit of the bare BeRP ball measurement. Next we used the value $k_c$ derived from MCNP6 simulations to solve for $l_c$ using Eq. \ref{eq:alpha}. This fully describes the behavior of the fissile core (BeRP ball). We next fit the remaining two free parameters, $l_r$ and $f$, to the TOFFEE distribution of the reflected configurations. 

\begin{figure}[htbp] 
	\centering
	\includegraphics[width=85mm]{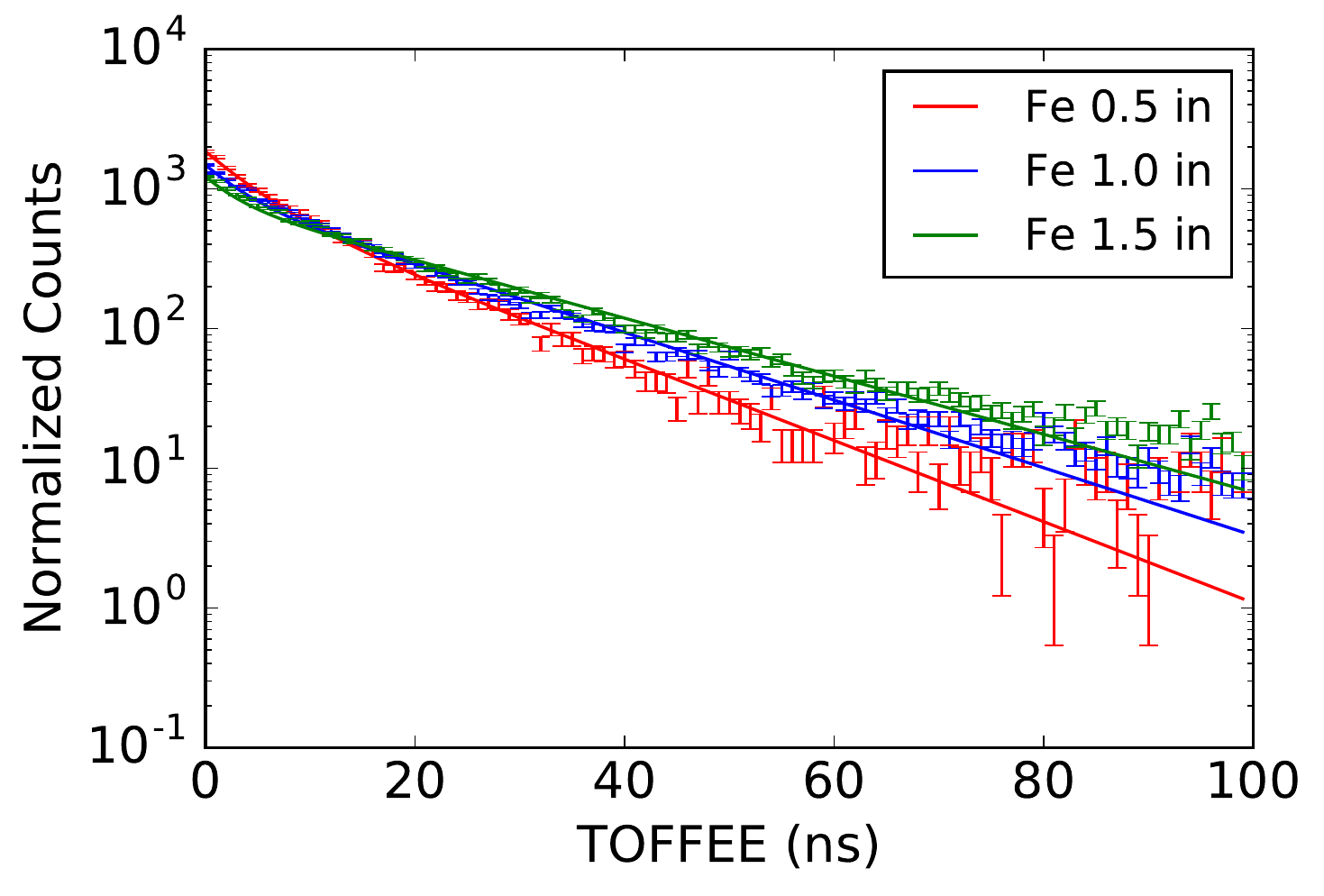}
	\caption{TOFFEE distributions of the measured iron configurations with corresponding double exponential fits from Eq. \ref{eq:Nc}.}
	\label{fig:double_exp_fit}
\end{figure}

The double exponential fit to the measured iron cases is shown in Figure \ref{fig:double_exp_fit}. In the first 60 ns time window the fit tracks quite well with the data, but undershoots the data at later times in the 80-100 ns window. Some of that is due to the lower statistic in that region which make it less important for the fit. There is also some effect of floor reflection that is not accounted for in the two-region point kinetics model and therefore missing from Eq. \ref{eq:Nc}. 

\begin{figure}[htbp] 
	\centering
	\includegraphics[width=85mm]{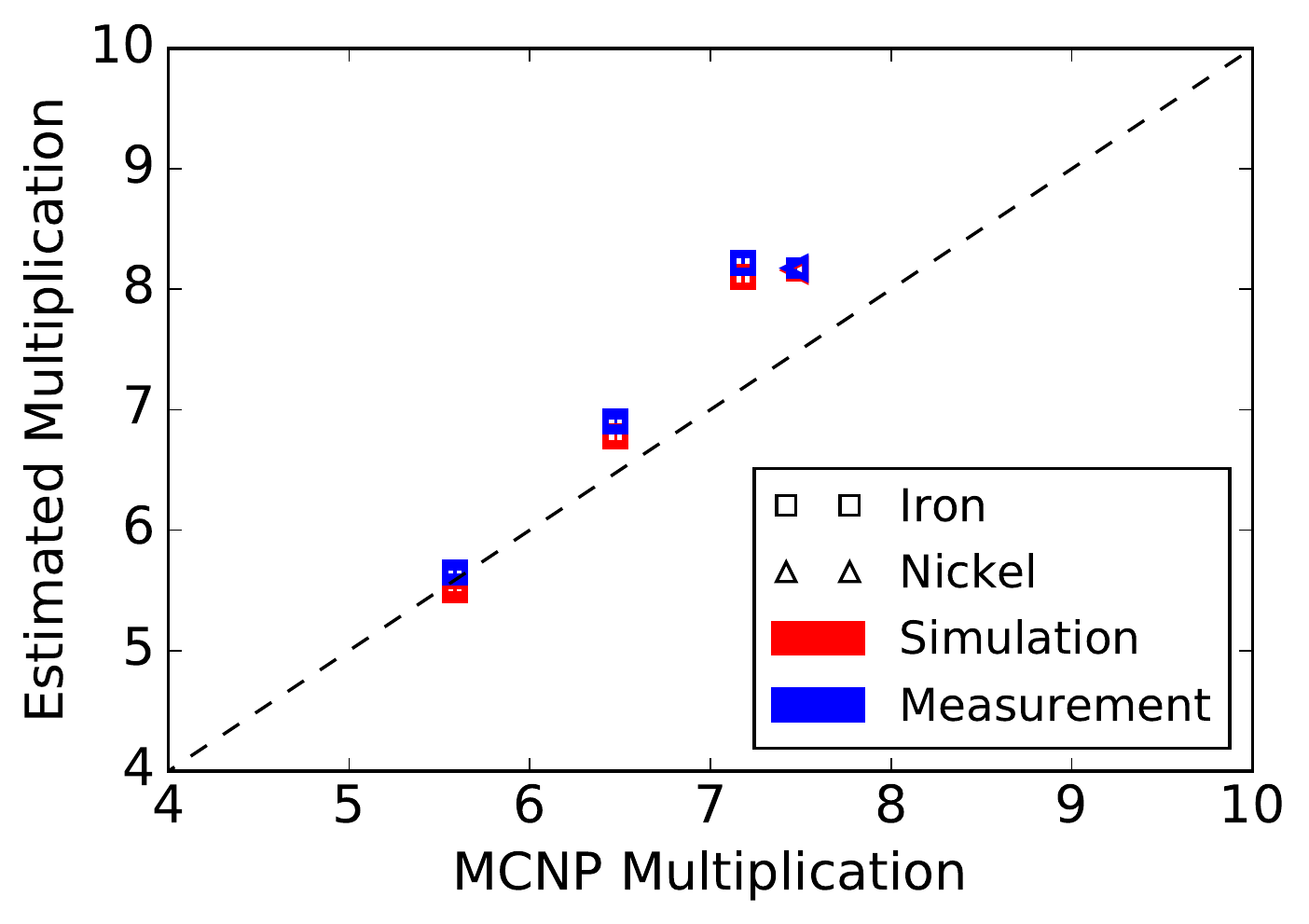}
	\caption{Comparison of the estimated multiplication of the measured and simulated TOFFEE distribution for the shielded configuration of the BeRP ball. The dashed line represent perfect agreement between the fit and the expectation from MCNP simulation.}
	\label{fig:mult_msr}
\end{figure}

The parameter $f$ is related to the total system $k$ by

\begin{align}
k = \frac{k_c}{(1-f)}.
\end{align}
Neutron multiplication is then calculated from Eq. \ref{eq:mult}. The comparison of this ``Estimated Multiplication" with the MCNP6 equivalent for the measured and simulated cases is shown in Figure \ref{fig:mult_msr}. As expected, there is convergence between the simulated and measured cases with increasing shell thickness. The average relative difference between estimated and expected multiplication was 10\%. 

\begin{figure}[htbp] 
	\centering
	\includegraphics[width=85mm]{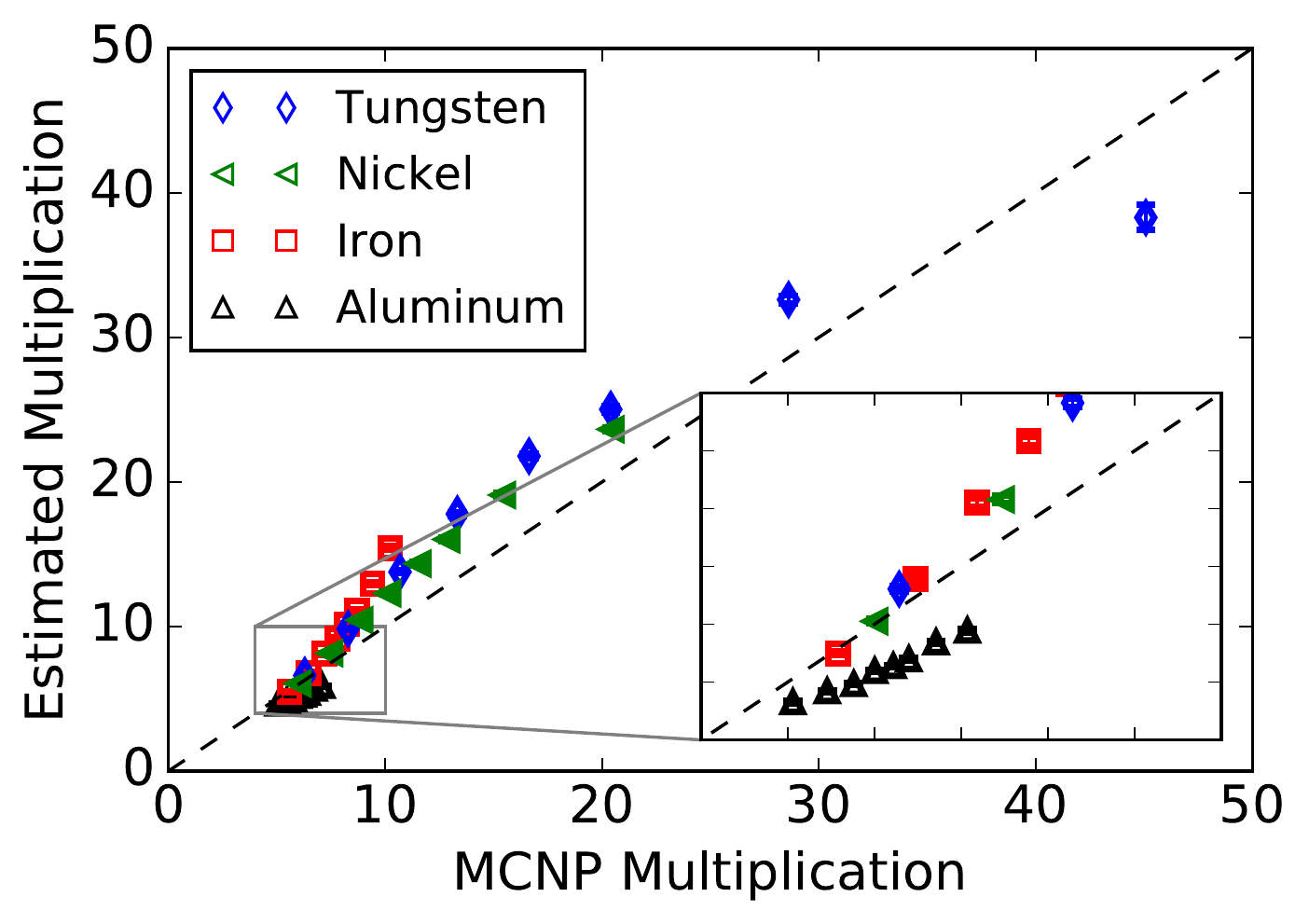}
	\caption{Estimated multiplication for simulated TOFFEE distributions of several configurations of shielded BeRP ball with different material types. }
	\label{fig:mult_sim}
\end{figure}

As before, we expanded on the measured cases with additional simulations, and the results are shown in Figure \ref{fig:mult_sim}. The estimated multiplications for all shielding materials have positive correlations with the MCNP multiplication, although the relationship is different between materials. The trend is superlinear for aluminum and sublinear for tungsten. Iron and nickel have a more linear trend. The average relative difference also varied from one material type to the next, with as little at 14\% for aluminum and as much at 22\% for iron. Unlike with the bare case correlation with leakage multiplication produced even worse agreement and the trends among the different materials remained the same. 

The discrepancies between materials are due to the assumptions in the two-region kinetic model. The model only allows for a neutron to either fission, or leak out of a region. There are no considerations for inelastic interaction, such as parasitic neutron capture, which differs among different materials. Neutron energy is also collapsed into one group, which works if neutron energy is not changing much. However, a neutron will on average lose more energy scattering in a low-Z aluminum, compared to a high-Z tungsten, due to scattering kinematics. As a result, a neutron reflected back into the fissile core from an aluminum reflector will have relatively greater probability of fission, due to the energy dependence of induced fission cross-section in Pu-239. This would in effect increase total multiplication, which may explain the underestimation of true multiplication for the aluminum reflected BeRP balls. 

\subsection{Shell Thickness and Material Type}

\begin{figure}[htbp] 
	\centering
	\includegraphics[width=85mm]{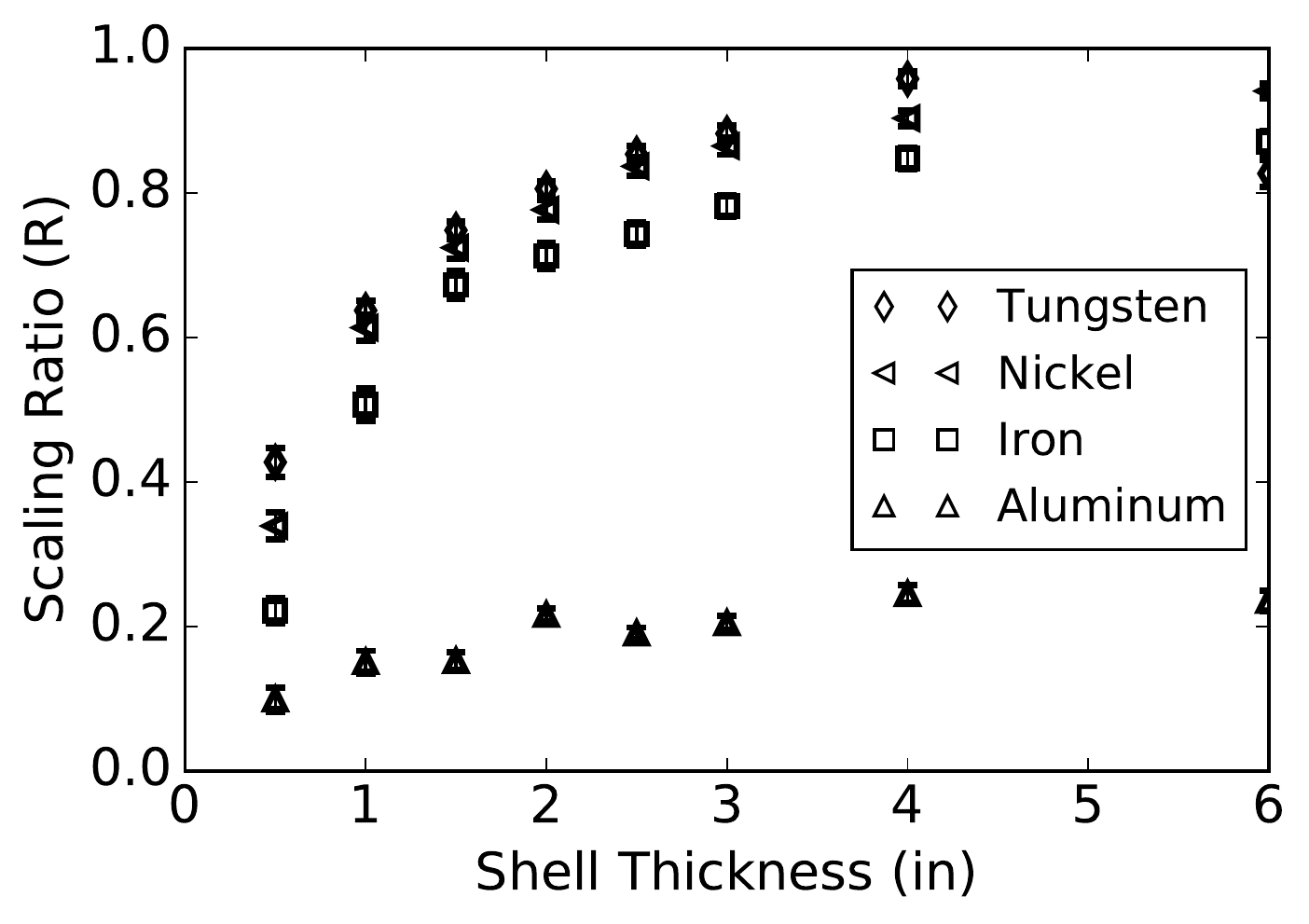}
	\caption{The scaling ratio from the fit of Eq. \ref{eq:Nc} to TOFFEE distributions of the BeRP ball with various reflector shell thicknesses.}
	\label{fig:ratio_sim}
\end{figure}

There are two derived quantities from the fitting procedure outlined in the previous Section \ref{sec:multiplication} which relate to shell thickness and material type. First, the scaling ratio asymptotically approaches unity with increasing shell thickness, as shown in Figure \ref{fig:ratio_sim}. As the amount of reflector material goes up, so does its effect on the TOFFEE distribution. Eventually this dominant reflector term collapses the double exponential fit into a single exponential. This suggests that it may be difficult to separate the effects fissile material mass and presence coupled reflector at sufficiently high amounts of said reflector. 

We quantified the amount of reflector material by calculating the effective areal density, which takes into account the hole in the shells. The integral of the aforementioned fits to TOFFEE distribution, defined as

\begin{align} \label{eq:Nc_integral}
\int_{0}^{\infty} N_c(t) dt = \frac{R - 1}{r_1} - \frac{R}{r_2}
\end{align} 
provided best linear correlation with the effective areal density. The results of linear least-squares regression for each material and all of them combined is shown in Table \ref{tab:integral_lin}. Each material has a unique linear correlation, but combined the regression shows a strong correlation coefficient of 0.9717. The deviation between materials is likely due to inelastic and other capture neutron interactions within each reflector, which we are not correcting for.

\begin{figure}[htbp] 
	\centering
	\includegraphics[width=85mm]{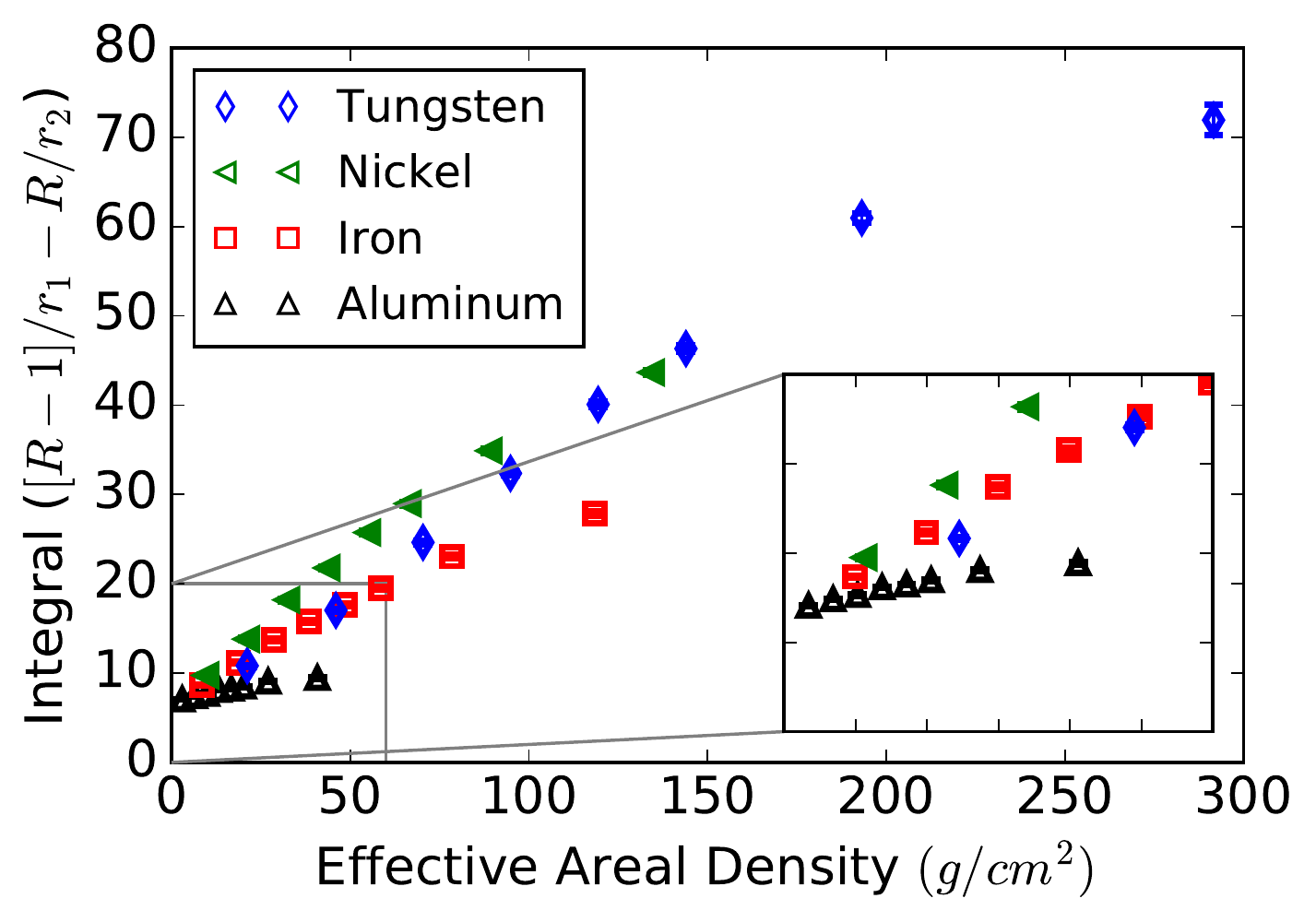}
	\caption{The integral of the fit of Eq. \ref{eq:Nc} to TOFFEE distributions of the reflected configurations of the BeRP ball and the effective areal density of each of the shells.}
	\label{fig:areal_sim}
\end{figure}

\begin{table}[htbp]
	\caption{Linear least-squared regression for the correlation between integral of the fit to TOFFEE distribution and effective areal density of the reflector material.}
	\centering
	\begin{tabular}{l| c c c}
		\hline
		\multirow{ 2}{*}{\textbf{reflector}} &  \multirow{ 2}{*}{slope}   &  \multirow{ 2}{*}{intercept}       & correlation  \\ 
		&    &  & coefficient\\
		\hline
		Aluminum  & 0.064  & 7.06 &  0.9748 \\ 
		Iron      & 0.175  & 8.46 &  0.9889 \\ 
		Nickel    & 0.274  & 9.10 &  0.9895 \\ 
		Tungsten  & 0.237  & 8.91 &  0.9815 \\
		All       & 0.251  & 6.55 &  0.9717 \\ \hline
	\end{tabular}
	\label{tab:integral_lin}
\end{table}
\clearpage
\section{Conclusions}

We introduced the TOFFEE distribution and its relationship to fission chain dynamics. We then fitted the positive side of this distribution, from 0 to 100 ns, to time dependent neutron population derived from point kinetics theory. A bare subcritical assembly is sufficiently described by a single exponential in Eq. \ref{eq:Nc_single}, and introduction of a reflector yields a double exponential shown in Eq \ref{eq:Nc}. 

We found that for a subset of bare BeRP ball masses, between 3 and 8 kg, the estimated alpha parameters and the expected alpha values are linearly correlated. A derived multiplication was calculated from the estimated alpha parameters  by assuming a known $l_c$ from MCNP6 simulations. This derived multiplication positively correlated with the leakage multiplication, with an average relative error of 10.6\%. 

The TOFFEE distributions from the reflected BeRP ball assemblies were fitted to the double exponential model from Eq. \ref{eq:Nc}. The derived multiplication from $f$ had a positive correlation with the expected multiplication for MCNP6, although the relationship varied between material types. Furthermore, we determined that the effective areal density of the reflectors was positively and linearly correlated with the integral of those same double exponential fits. It is conceivable that with knowledge of either shell thickness or material composition it would be possible to determine the other property.

\section*{Acknowledgments} 

This material is based upon work supported by the U.S. Department of Homeland Security under Grant Award Number, 2012-DN-130-NF0001. The views and conclusions contained in this document are those of the authors and should not be interpreted as representing the official policies, either expressed or implied, of the U.S. Department of Homeland Security.

Sandia National Laboratories is a multimission laboratory managed and operated by National Technology and Engineering Solutions of Sandia, LLC., a wholly owned subsidiary of Honeywell International, Inc., for the U.S. Department of Energy’s National Nuclear Security Administration under contract DE-NA-0003525.

This work was funded in-part by the Consortium for Verification Technology under Department of Energy National Nuclear Security Administration award number DE-NA0002534.

\clearpage
\section*{References}

\bibliography{mybibfile}

\end{document}